\documentclass[fleqn,usenatbib]{mnras}

\usepackage{newtxtext,newtxmath}

\usepackage[T1]{fontenc}

\DeclareRobustCommand{\VAN}[3]{#2}
\let\VANthebibliography\thebibliography
\def\thebibliography{\DeclareRobustCommand{\VAN}[3]{##3}\VANthebibliography}

\usepackage{graphicx}	
\usepackage{amsmath}	
\usepackage{amssymb}	
\usepackage[table]{xcolor}

\title[A Wide-Field View on Multiple Populations]{A Wide-Field View on Multiple Stellar Populations in 28 Milky Way Globular Clusters}

\author[E. I. Leitinger et al.]{
E. Leitinger,$^{1,2}$\thanks{E-mail:  ellen.leitinger@uq.net.au}
H. Baumgardt,$^{1}$
I. Cabrera-Ziri$^{3}$
M. Hilker$^{2}$
and E. Pancino$^{4,5}$
\\
$^{1}$School of Mathematics and Physics, The University of Queensland, St. Lucia, QLD, 4072, Australia\\
$^{2}$European Southern Observatory, Karl-Schwarzschild-Str. 2, 85748 Garching, Germany\\
$^{3}$Astronomisches Rechen-Institut, Zentrum f\"ur Astronomie der Universit\"at Heidelberg, M\"onchhofstra{\ss}e 12-14, D-69120 Heidelberg, Germany\\
$^{4}$INAF – Osservatorio Astrofisico di Arcetri, Largo Enrico Fermi 5, I-50125 Firenze, Italy\\
$^{5}$Space Science Data Center, ASI, via del Politecnico snc, I-00133 Roma, Italy
}

\date{Accepted 19 December 2022}

\pubyear{2023}

\begin{document}
\label{firstpage}
\pagerange{\pageref{firstpage}--\pageref{lastpage}}
\maketitle

\begin{abstract}
The majority of Galactic globular clusters (GCs) contain multiple stellar populations displaying specific chemical abundance variations. In particular, GCs generally contain a `primordial' population with abundances similar to field stars, along with an `enriched' population exhibiting light element anomalies. In this paper we present a homogeneous and wide-view analysis of multiple stellar populations in 28 Galactic GCs. By using a combination of \textit{HST} photometry together with wide-field, ground-based photometry we are able to analyse between 84\% and 99\% of all stars in each cluster. For each GC, we classify stars into separate sub-populations using the well-established $C_{\rm{UBI}}$ colour index, and investigate the spatial distributions of these populations.
Our results show that dynamically young GCs can contain either centrally concentrated enriched or primordial populations, \textit{or} no centrally concentrated population. Dynamically old GCs show fully mixed populations as expected. The existence of clusters born with centrally concentrated primordial (and homogeneously mixed) populations exacerbates the mass-budget problem facing many cluster formation scenarios. The diversity in these results also highlights the need for additional theories that can account for the wide variety of initial conditions that we find. We finally investigate the enriched star fraction as a function of different global parameters in our GC sample, using also data for young and low-mass clusters from the Small- and Large Magellanic Clouds and confirm earlier results that the enriched star fraction strongly correlates with the initial mass of a cluster.
\end{abstract}

\begin{keywords}
(Galaxy:) globular clusters: general -- Stars: abundances -- (stars:) Hertzsprung–Russell and colour–magnitude diagrams -- stars: kinematics and dynamics -- Galaxy: evolution
\end{keywords}



\section{Introduction}

Most Galactic globular clusters (GCs) contain multiple stellar populations (MPs), distinguished by star-to-star variations in light element abundances that are not explained by simple stellar evolution. Stars are determined as `primordial' (usually as P1) if their elemental abundances are similar to the surrounding field stars of the cluster, and `enriched' (P2) if they demonstrate an enhancement in some light elements (e.g. He, N, Na and Al), but a depletion in others (e.g. C, O and sometimes Mg) in comparison to P1 \citep{2012gratton,2016charbonnel,2018bastian}. However, heavier element variations such as Fe only are present in a minority of clusters  \citep{2009carretta,2012willman,2022bastian}. The formation history of GCs necessary to produce MPs is a matter of ongoing debate \citep{2018forbes,2019gratton,2020cassisi}. We know that the observed abundance patterns are compatible with the chemistry of the CNO-cycle (and hot subcycles) and that this happens mostly in massive stars or in the H-burning shells of red giants, which leads to the theory that stellar formation of the enriched populations is fuelled by GC internal processes.

An important piece of information regarding the formation history of MPs in GCs is the spatial distribution of the stars in each population. If a cluster has not undergone significant dynamical mixing during its lifetime, we can assume it still maintains its initial spatial configurations. If we then observe that one stellar population is located primarily within the centre of such a cluster, we can assume this was the initial configuration of the stars during cluster formation. The analysis presented in this paper focuses in part on the spatial distribution of the MPs, which serves as a way to test the validity of the current processes theorised to describe cluster formation.

One such process is the AGB scenario, first proposed by \cite{1981cottrell}, in which first generation (P1) AGB stars expel enriched material by stellar winds, which accumulates in the center of the cluster and mixes with primordial material to spark a second event of star formation - creating P2 stars. However, for clusters in which the P2 population is equal to, or more massive than, the P1 population, the AGB scenario encounters a `mass budget' problem since, assuming a standard stellar mass function, the enriched material created from P1 stars is not sufficient to create the P2 stars we observe in some clusters \citep[e.g.][]{2006prantzos,cz15}. An implication of the AGB scenario is that an enriched star formation event occurring in the center of the cluster will lead to centrally concentrated P2 stars. 

Another formation process involves enrichment due to Super Massive Stars (SMS) \citep{2014denissenkov,2018gieles} that form due to runaway collisions in the early stages of cluster formation and reside in the center of a cluster, providing a `conveyer belt' of enriched material with different He fractions. This theory can overcome the mass budget problem as the continuous stellar collisions provide additional Hydrogen, which constantly rejuvinates the SMS. In this theory, P2 star formation occurs in the regions surrounding the SMSs. 

Fast rotating massive stars were proposed by \cite{2007decressin,2007charbonnel} to account for the observed chemical inhomogeneities, as massive stars create the required enriched material for additional star formation events, while the fast rotation brings the material to the surface of the star and ejects it. In this scenario, secondary star formation events occur in the region surrounding the fast rotating massive stars after the enriched material is diluted by left over primordial gas.

Finally, massive interacting binaries have been suggested as a probable cause for the chemical enrichment found in MPs of GCs by \cite{2009demink} and \cite{2022renzini}. \cite{2022renzini} theorised that above a certain critical mass threshold, massive stars skip the supernova stage and instead implode into black holes, therefore ensuring the remaining stars in the cluster do not contain an abundance spread in Fe. As the centers of GCs are much denser than the outer regions, binary stars are expected to be destroyed or ejected at a higher rate in the center than they do in the outer regions due to increased collisions. \cite{2015lucatello} discovered a higher fraction of binaries within the P1 population, as opposed to the P2 population in 10 Galactic GCs, which seems to support theories that assume P2 stars are centrally concentrated. \\

\cite{2019dalessandro} studied the radial distribution of 20 Galactic GCs as a function of the age/relaxation time fraction (hereby referred to as `dynamical age') using HST photometry and N-body model simulations. They found that clusters with low dynamical ages preferentially contain centrally concentrated P2 populations. It is expected that clusters with lower dynamical ages have not undergone much dynamical mixing in their lifetime and are therefore still exhibiting properties close to their initial conditions. Clusters with higher dynamical ages were found to have spatially blended multiple stellar populations, in agreement with the idea that these clusters have undergone significant dynamical mixing. The results found by \cite{2019dalessandro} provide observational evidence for formation theories in which enriched populations are formed within the center of the cluster. In their review, \cite{2018bastian} concluded that GCs might not have homogeneous histories, suggesting instead that MPs can be formed through a variety of individual scenarios. In this case we could assume that the scenarios mentioned above are responsible for clusters with centrally concentrated P2 stars. However, if a cluster contains centrally concentrated P1 stars, there are no current theories to explain this.\\

In this work, we study a diverse sample of 28 Galactic GCs in order to provide a comprehensive insight into the various possibilities of cluster properties. Large scale photometric analyses have been performed on Galactic GCs by \cite{2013monelli,2017milone,stetsonphotometry}, revealing intriguing scaling relations that may help us understand the origin of MPs. So far, combined space- and ground-based photometry for the purpose of obtaining a thorough spatial analysis of MPs and their characteristics exists only for a small number of clusters. We used both space-based and ground-based photometry to perform a homogeneous analysis of the wide-field spatial extent of a large sample of GCs, using the well-established color combination $\rm{C_{UBI}}$ and chromosome map methods in order to separate the MPs. In this paper we categorise MPs in space- and ground-based photometry separately, before combining the results to investigate correlations in terms of spatial distributions, enriched star fractions and global properties. We also compare our results with theoretical data and combine the Galactic GCs with Local Group GCs to further investigate trends.

\section{Observational Data}
\label{sec:obsdata}

The photometric catalogues used in this work include the wide-field ground-based Johnson-Cousins \textit{UBVRI} photometric data provided by \cite{stetsonphotometry}, along with the space-based HST UV Globular Cluster Survey data (`HUGS') \citep{hugs2, hugs1} with photometry obtained through UV/blue and WFC3/UVIS filters. For this first project we will focus our analysis of multiple stellar populations only on the RGB stars of these catalogues, combining both the HST and ground-based photometry in order to observe a wide-field view of each cluster, covering at least 84\% of the stars. The \cite{stetsonphotometry} photometric catalogue includes 48 GCs and the HUGS survey includes 57 GCs, but only 32 of these clusters overlap and exist in both catalogues. Of these 32 clusters, we successfully classified distinct MPs in 28 of them. We excluded clusters from our sample if they contained too few RGB stars after removing non-members and performing photometric cleaning, or if the classification of cluster stars into different sub-populations was inconclusive. The ground-based catalogues cover almost the full extent of each cluster, but cluster centers have much higher stellar densities than the outer regions, causing blending to affect the photometry of stars close to the center. This is where using HST photometry for the inner regions of clusters has an advantage, as crowding is less of an issue with space-based photometry. 

In this section we detail the steps taken to remove non-members, non-RGB stars and bad photometry from each photometric catalogue before separating the multiple stellar populations in Section \ref{sec:popsplit} and characterising the cluster properties in Section \ref{sec:results}.

Both the ground-based and HST catalogues encountered issues with different types of incompletenesses. In areas of the observed fields where either no stars were measured in a relevant filter, or the photometry was too poor to be usable, we could not reliably make assumptions about the properties of stars in that area. We calculated completeness fractions for the remaining stars so that we account for the stars that were missed. We describe the spatial incompleteness in Section \ref{sec:spatial}, the photometric incompleteness in Section \ref{sec:phot_incompleteness} and the surface density incompleteness in Section \ref{sec:magnitude_completeness}.

\begin{figure}
    \centering
    \includegraphics[width=\columnwidth]{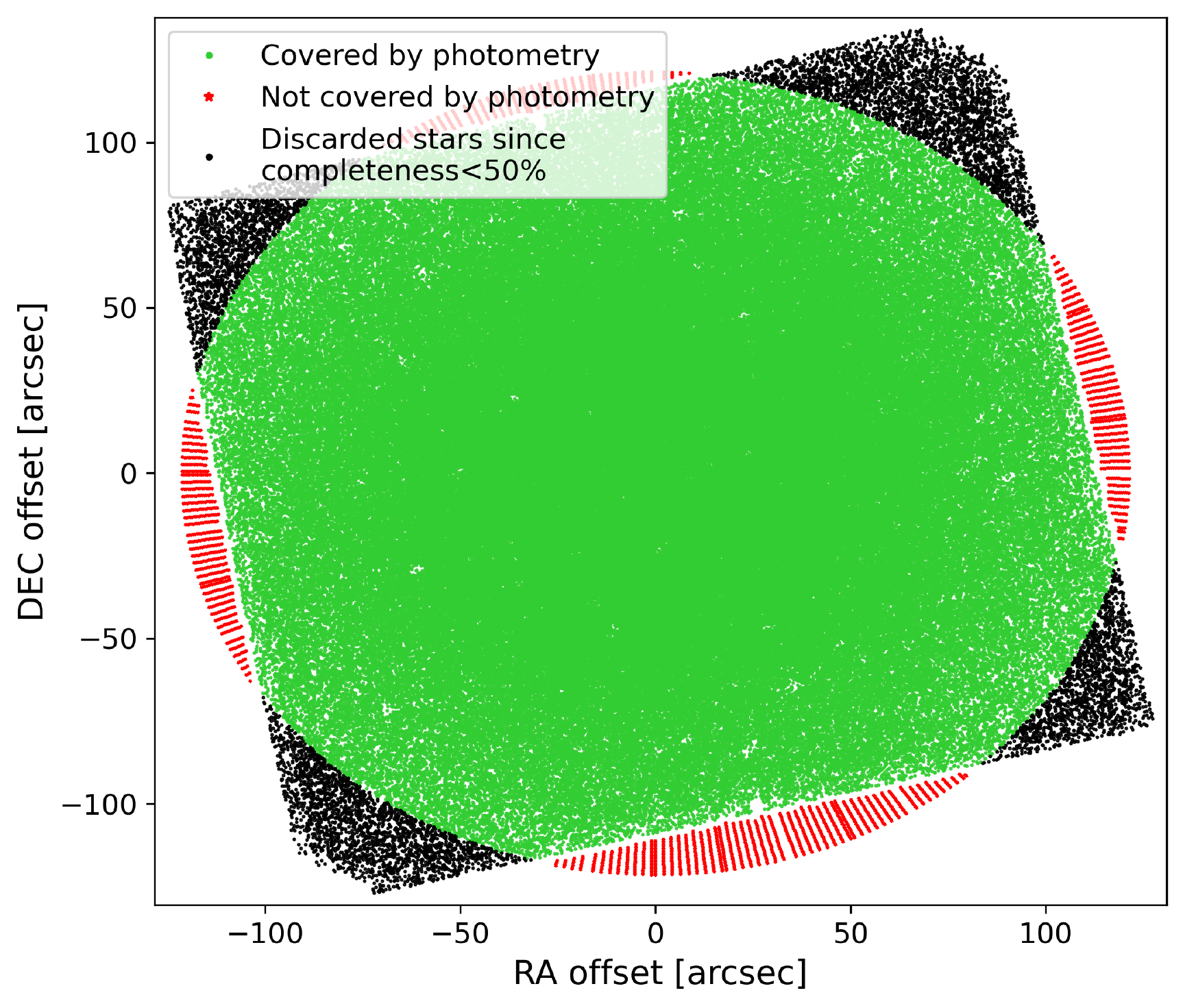}
    \caption{Spatial completeness of the HST photometry for the globular cluster NGC 5024. Artificial test stars are shown in red (green), if our test indicated they fall outside (inside) the area covered by photometry. Stars shown in black are real stars located in regions that fall below 50\% completeness.}
    \label{fig:spatial_incompleteness}
\end{figure}

\subsection{Spatial Completeness Correction}
\label{sec:spatial}

Our first step in processing both the ground-based and HST catalogues was determining the spatial completeness fraction $f_S$ of each catalogue independently. Using the original catalogues for both the ground-based and HST photometry, the spatial position of each star in right ascension and declination were calculated as an offset from the cluster center. The data was not cleaned for stars without measured photometry, defined as $\rm{mag} < 0$ for HST and $\rm{mag} > 99$ for the ground-based photometry, since these entries in the catalogues still indicated the presence of a star.

We distributed a series of concentric rings spaced by $1.0''$ in distance around the cluster centers and distributed 360 artificial points evenly spaced by 1 degree along each ring. For each of these artificial points, we determined the distance between the point and its nearest star, from the surrounding stars in our photometry. A point was considered to be covered by the photometry if the minimum distance was less than a tolerance distance - usually close to 1 arcsec, but otherwise dependent on the cluster. This method has the flexibility to be able to account for arbitrary field geometries, including large gaps within the field.

The spatial completeness $f_S$ of each annulus was set equal to the fraction of points that were covered by photometry in the field: $ f_S = \frac{N_{\rm{in}}}{N_{\rm{total}}}$, where $N_{\rm{in}}$ is the number of points inside the observed field and $N_{\rm{total}} = 360$, the total number of points for that annulus. We discarded photometry outside the radius in which the spatial completeness drops below 50\%, shown as black points in Figure \ref{fig:spatial_incompleteness}, using NGC 5024 as an example. Surviving stars were assigned a spatial completeness fraction ($0.5 \leq f_S \leq 1.0$), based on the completeness of the annulus they were located within. The HST and ground-based data was combined without allowing spatial gaps in the field by ensuring the ground-based data begins at the same radius at which the HST data ends for all clusters.

\subsection{Differential Reddening Correction}
\label{sec:reddening}
\begin{figure}
    \centering
    \includegraphics[width=\columnwidth]{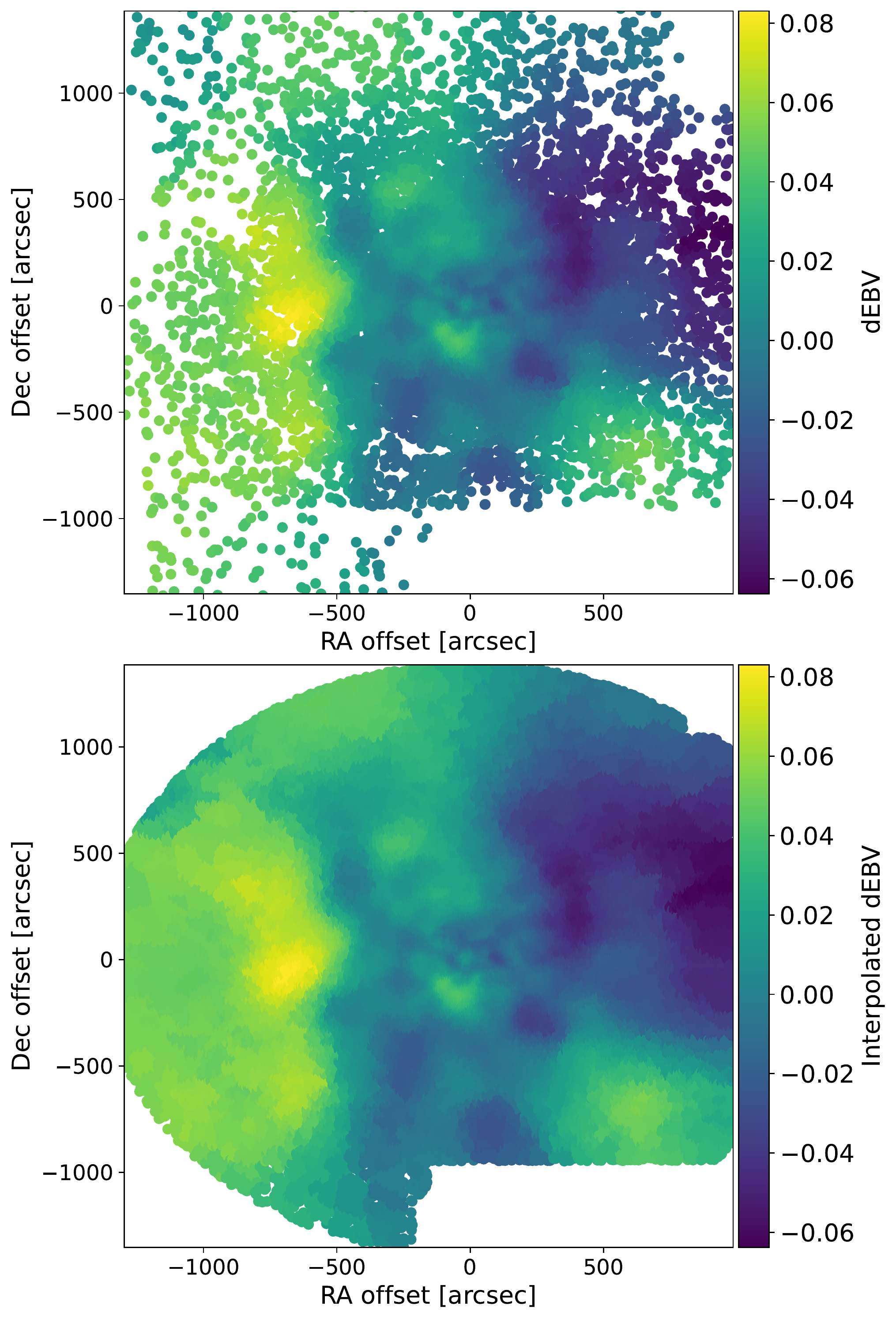}
    \caption{\textit{Top panel:} Differential reddening map of NGC 6121. \textit{Bottom panel:} Interpolation of the reddening map onto the ground-based photometry after spatial completeness correction in order to assign individual values of dEBV based on the nearest-neighbour in the top panel.}
    \label{fig:reddening}
\end{figure}

To compute differential reddening maps, we used a method similar to other methods employed in the literature \citep[e.g.,][]{milone12}, which will be described in detail in a forthcoming publication (Pancino et al., in preparation). We used the ground-based photometry by \citet{stetsonphotometry}, selecting stars with photometric errors lower than 0.3~mag in $BVI$, $\chi < 3$, and $|$sharp$|$ < 0.5. We computed a fiducial line as the median ridge line of the main sequence of each cluster, down to about 2--4 magnitudes below the turnoff point. We selected stars not further than the 5 and 95\% percentiles from the fiducial line in the three color planes $V$,$B$--$V$; $V$,$V$--$I$; and $V$,$B$--$I$. This allowed us to remove a large fraction of contaminating field stars. The color difference of each selected star from the reference line was computed in the three planes along the reddening line, assuming R$_V$ = 3.1 and using \citet{dean1978} to compute the reddening line direction in each plane. We then rescaled these raw color differences and combined them into one single estimate of $\Delta$E(B--V) for each star. To disentangle photometric errors and other effects from the actual differential reddening signal, we smoothed these maps in right ascension and declination. by replacing the $\Delta$E(B--V) of each star with the median of its $k$ neighbors, with $k$ ranging from 50 to 300 (typically in the range 150-200) depending on the cluster. This also allowed us to compute an uncertainty for each differential reddening estimate as the median absolute deviation of the values for the $k$ neighbours.

To correct the ground-based photometry, the reddening map was interpolated for each star in both the HST and ground-based catalogues, as shown in the bottom panel of Figure \ref{fig:reddening}. We used the standard ratio of absolute to selective extinction of $R_V = 3.1$, with the exception of NGC 6121, for which the value of $R_V = 3.76 \pm 0.07$ was used as suggested by \citet{hendricks2012}. Magnitude corrections for the ground-based $U$ and $B$ bands were applied using extinction ratios according to \citet{cardelli1989}, while the $R$ and $I$ bands were corrected according to \citet{dean1978}. Similarly for the HST photometry, differential reddening was corrected for the $F275W, F336W, F438W$ and $F814W$ bands using extinction ratios from the SVO Filter Profile Service \citep{SVO1,SVO2}.

\subsection{Photometric Quality Indicators}
\label{sec:quality_indicators}

\begin{figure}
    \centering
    \includegraphics[width=\columnwidth]{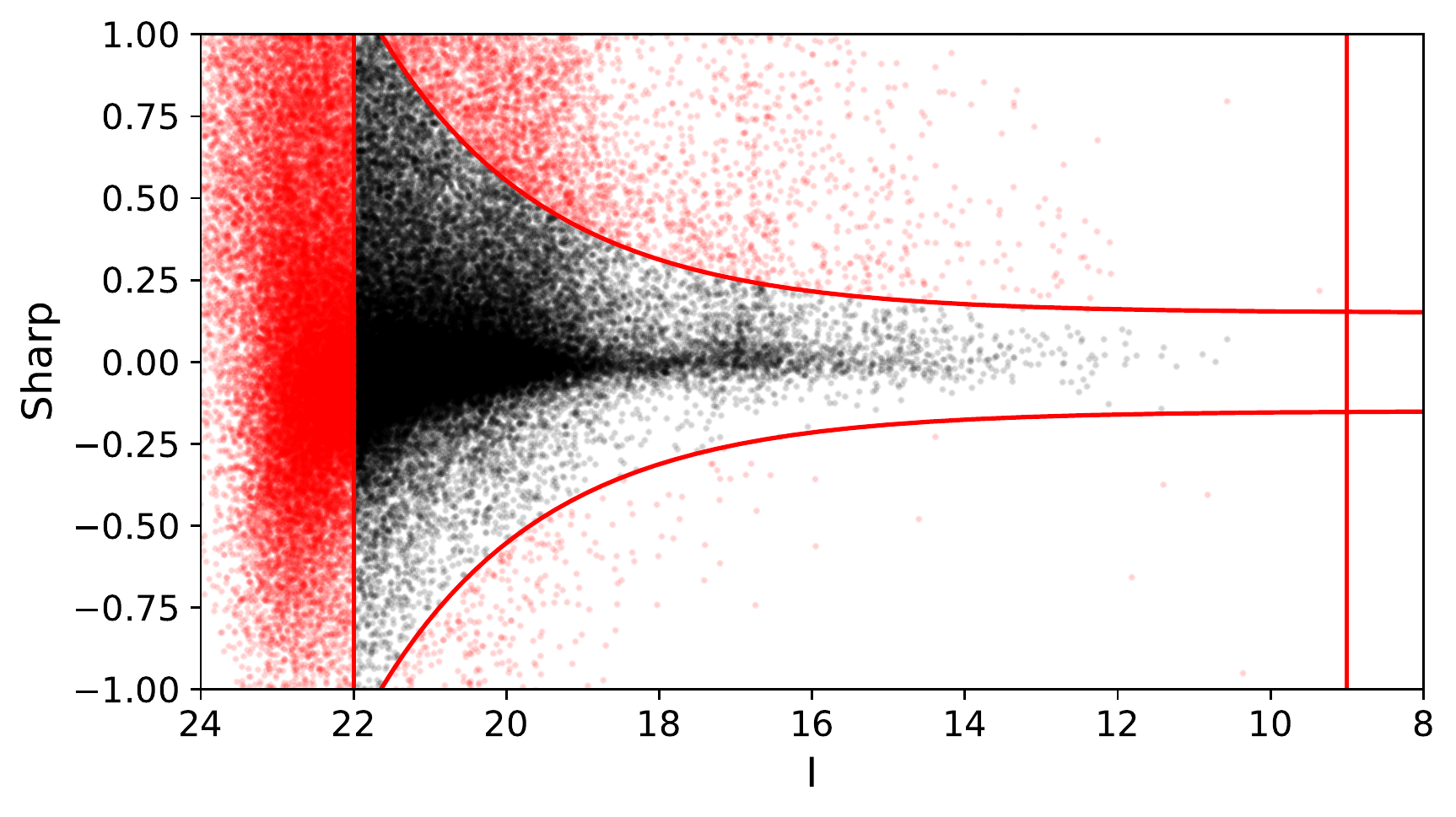}
    \caption{Sharp parameter cuts for the ground-based photometry of NGC 5024. Black points represent stars that survived the cut, red points were removed. The two red vertical lines represent rough limits in magnitudes to isolate the RGB. An `envelope' function in red encloses stars with large enough photometric quality, as defined by Equation \ref{eq:sharp}.}
    \label{fig:sharp}
\end{figure}

\begin{figure}
    \centering
    \includegraphics[width=\columnwidth]{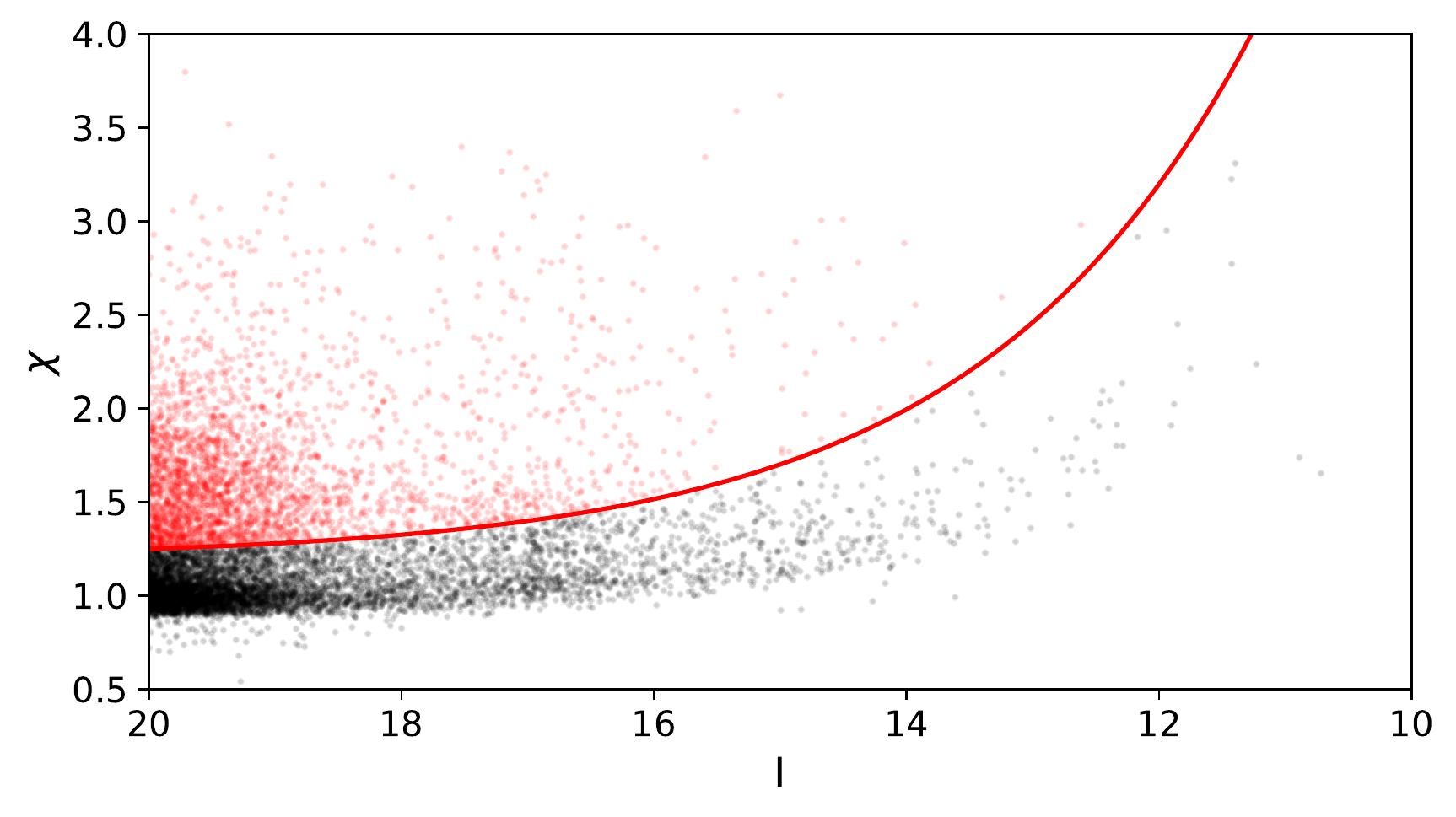}
    \caption{$\chi$ parameter cuts performed only on the ground-based photometry. Black points represent the stars of NGC 5024 that survived the \textit{sharp} cuts of Figure \ref{fig:sharp}, while red points were removed. All stars beneath the red line, defined by Equation \ref{eq:chi} are kept.}
    \label{fig:chi}
\end{figure}

We removed stars with less reliable photometry by using different quality indicators based on the available parameters provided by the HST and ground-based catalogues. For the ground-based photometry, we implemented quality cuts based on magnitude errors and the $\chi$ and \textit{sharp} parameters described in the work of \cite{stetsonharris1988}. For the HST photometry we implemented cuts in \textit{sharp} while also using the membership probability and quality-fit parameters (QFIT) for each star provided by \cite{hugs1}. For the ground-based photometry, the \textit{U,B,V} and \textit{I} bands with associated errors $>9$ mag  were cut. For the HST photometry, using the same constraints as \cite{2019dalessandro}, stars belonging to the cluster were selected using membership probability $>75\%$ and $QFIT > 0.9$ in each of the $F336W, F438W, F606W$ and $F814W$ bands.

For both photometry sets, cuts were made based on the \textit{sharp} values following a method similar to \cite{2003Stetson}, but replacing the $-1 \geq \mbox{sharp} \geq 1$ criterion with an `envelope' function. We defined an exponential function above and below the bulk of the values to remove stars with sharp values too far from the mean:
\begin{equation}
    |\mbox{sharp}| < 0.15 + exp\left(\frac{mag - 22}{1.5}\right),
    \label{eq:sharp}
\end{equation}

where $mag = I$ for the ground-based photometry and $mag = F814W$ for HST. Figure \ref{fig:sharp} shows the cut for ground-based photometry in which stars enclosed within the envelope are kept.\\

For the ground-based photometry we also used the $\chi$ parameter, which determines the observed vs. expected pixel-to-pixel scatter. By adapting the method from \cite{2003Stetson}, a function was applied to remove outliers:
\begin{equation}
    \chi < 1.2 + {2\times10^{\left(-0.2(I - 12)\right)}} .
    \label{eq:chi}
\end{equation} 
Stars which met the criterium are shown in black in Figure \ref{fig:chi}, while stars in red were rejected. 

\subsection{Proper Motion Cleaning}
\begin{figure}
    \centering
    \includegraphics[width=\columnwidth]{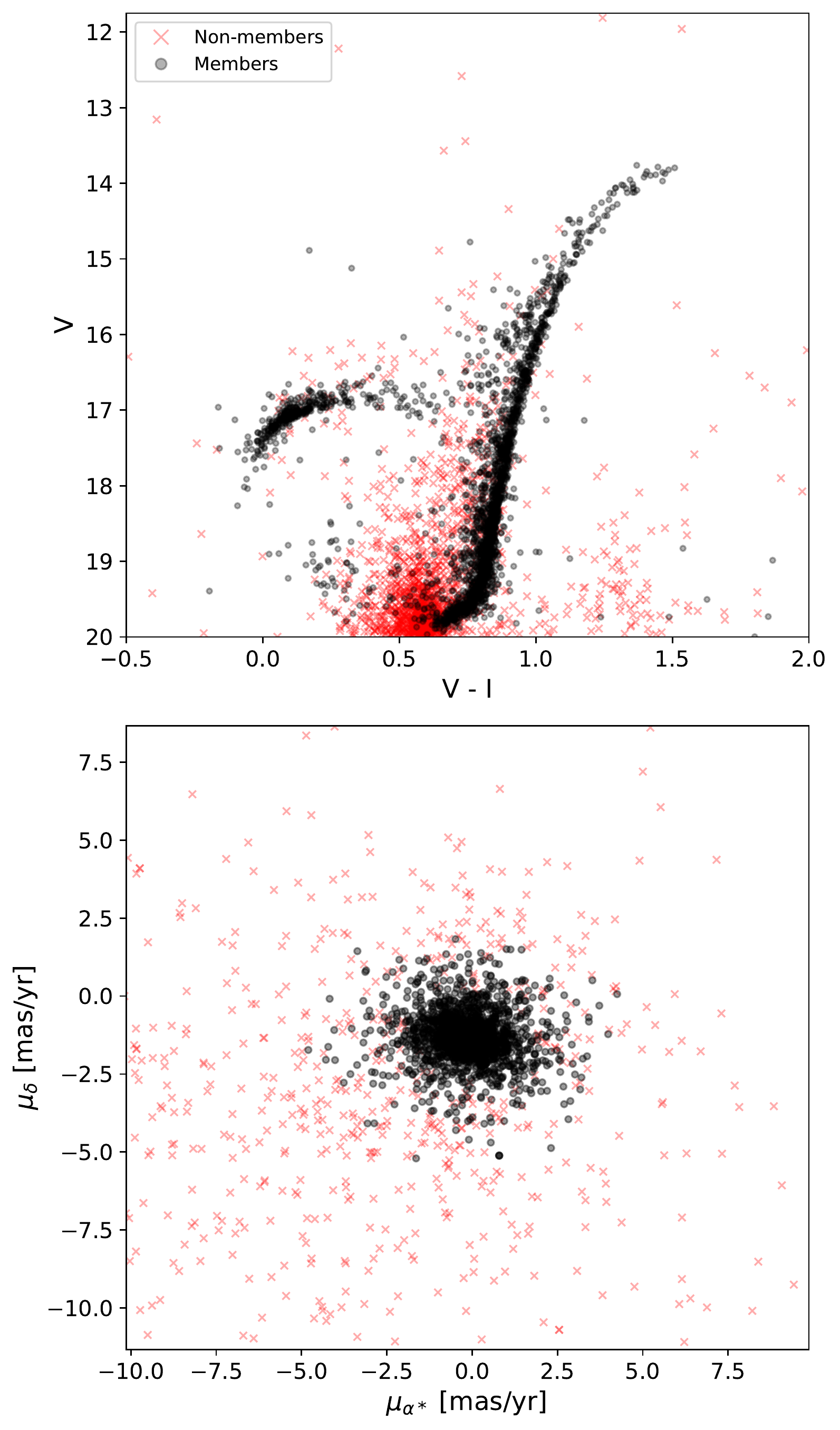}
    \caption{Ground-based photometry for NGC 5024, demonstrating the effect of proper motion cleaning. \textit{Upper panel}: CMD of stars above the approximate MS turn-off, with accepted stars in black and rejected stars in red. \textit{Lower panel}: The proper motion distributions of stars matched with Gaia EDR3, divided into members (black) and non-members (red). }
    \label{fig:pmcleaned}
\end{figure}
The HST photometry includes a membership probability parameter \cite[see][]{hugs1} to help discard stars that do not belong to the cluster, as discussed in Section \ref{sec:quality_indicators}. Determining the true members of a cluster for the ground-based photometry was done using proper motions of the stars after cross-matching with the Gaia DR3 catalogue \citep{edr32016b,edr32020a}. This catalogue is comprehensive in scale, but has difficulties with incompleteness in the center of clusters and lower accuracy due to the high stellar crowding \citep{clusterdatabase}.

In order to enforce an equivalent MS turn-off limit between all catalogues, we first located the MS turn-off in the Gaia G band and applied a cut exactly at this magnitude to isolate the RGB stars. This was a precaution against matching faint stars from one catalogue to bright stars in another catalogue (for stars in close proximity to each other). We then cross-matched between the Gaia, HST and ground-based catalogues within a $0.5 ''$ tolerance and determined the equivalent MS turn-off in the HST and ground-based catalogues. We isolated the RGB stars in each catalogue using the equivalent MS turn-off limits found from this process.

Proper motion cleaning was only performed on the ground-based photometry outside the HST footprint due to the aforementioned high stellar densities in the center of the clusters. The stars matched with the Gaia catalogue were then proper motion cleaned using a $\chi^2$ test, defined in Equation \ref{eq:pmclean}, using both the right ascension $\mu_{\alpha*}$ and declination $\mu_\delta$ proper motion components and corresponding errors. The cluster proper motion values ($\mu_{\alpha*,cluster}$ and $\mu_{\delta,cluster}$) were taken from \cite{clusterdatabase}. We include a proper motion error of $0.2$ mas/yr to account for both the internal velocity dispersion of the cluster and any proper motion errors that may be underestimated.
\begin{equation}
    \chi^2 = \frac{(\mu_{\alpha*,cluster} - \mu_{\alpha*})^2}{(\mu_{\alpha*,err})^2 + 0.2 \rm{[mas/yr]} ^2} + \frac{(\mu_{\delta,cluster} - \mu_{\delta})^2}{(\mu_{\delta,err})^2 + 0.2 \rm{[mas/yr]} ^2}
    \label{eq:pmclean}
\end{equation}
The cut-off limit for the $\chi^2$ value was slightly varied for each cluster, depending on the background stellar density and how clearly the cluster motion was distinguishable from the background. In order to limit the effect of large errors allowing non-members to pass, we implemented an error tolerance relative to the proper motion of the cluster. The resulting cluster member stars are shown in black in both panels of Figure \ref{fig:pmcleaned}, while rejected stars are shown in red. The ground-based stars within $100 ''$ of the cluster center were added to the confirmed cluster member stars for the photometric cleaning in Section \ref{sec:photclean}. We did this as these inner ground-based stars assisted with photometric cleaning and were removed anyway once the HST and ground-based photometry were combined.

\subsection{Photometric Cleaning}
\label{sec:photclean}

\begin{figure}
    \centering
    \includegraphics[width=\columnwidth]{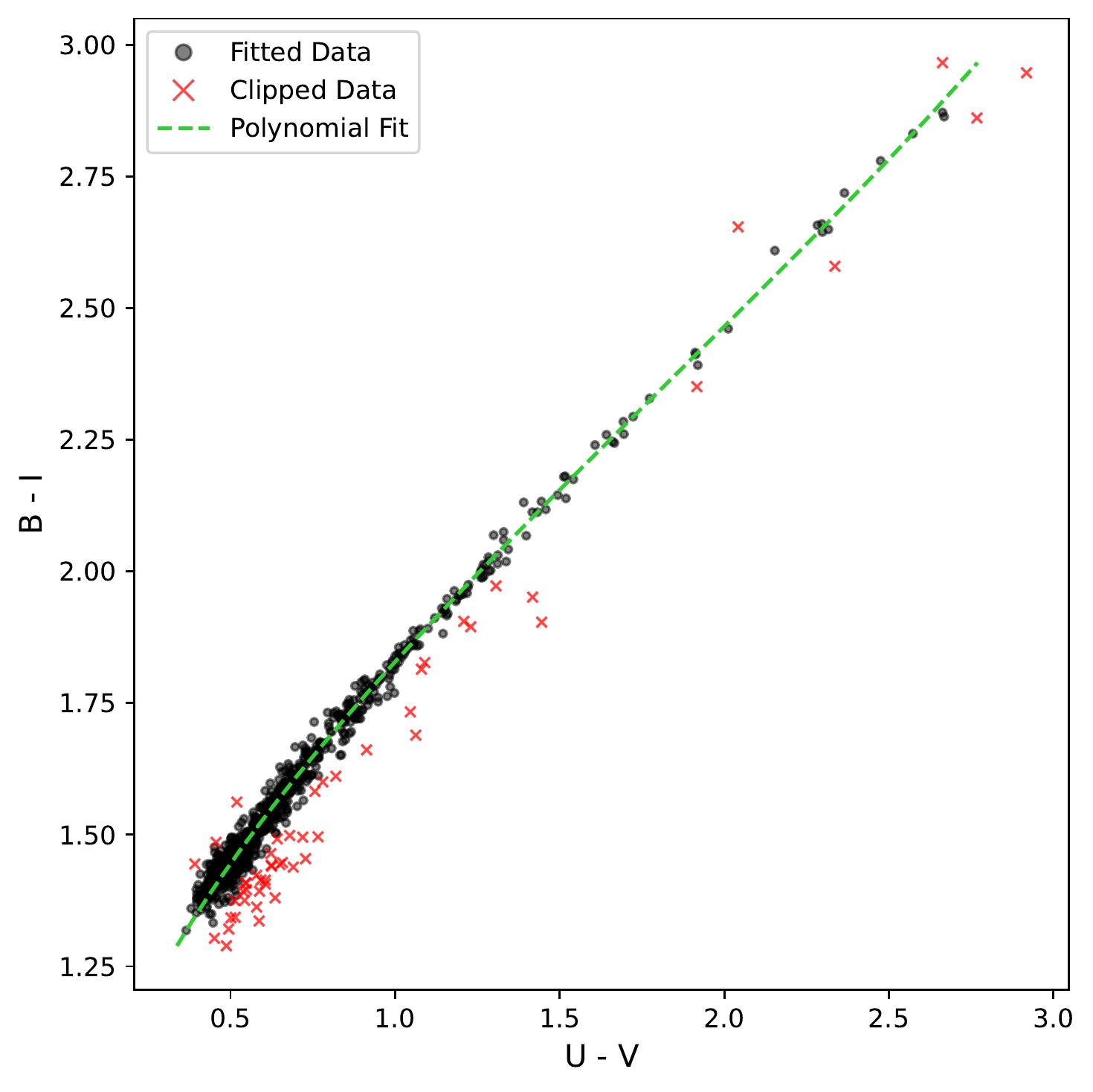}
    \caption{Polynomial fitting of RGB stars in colour-colour combinations $U - V$ vs $B - I$ for the ground-based photometry of NGC 5024. The line of best fit for the RGB stars is in green, cluster members are in black and non-members removed via the $N-\sigma$ clipping method are in red.}
    \label{fig:UVBI}
\end{figure}
\begin{figure*}
    \centering
    \includegraphics[width=\textwidth]{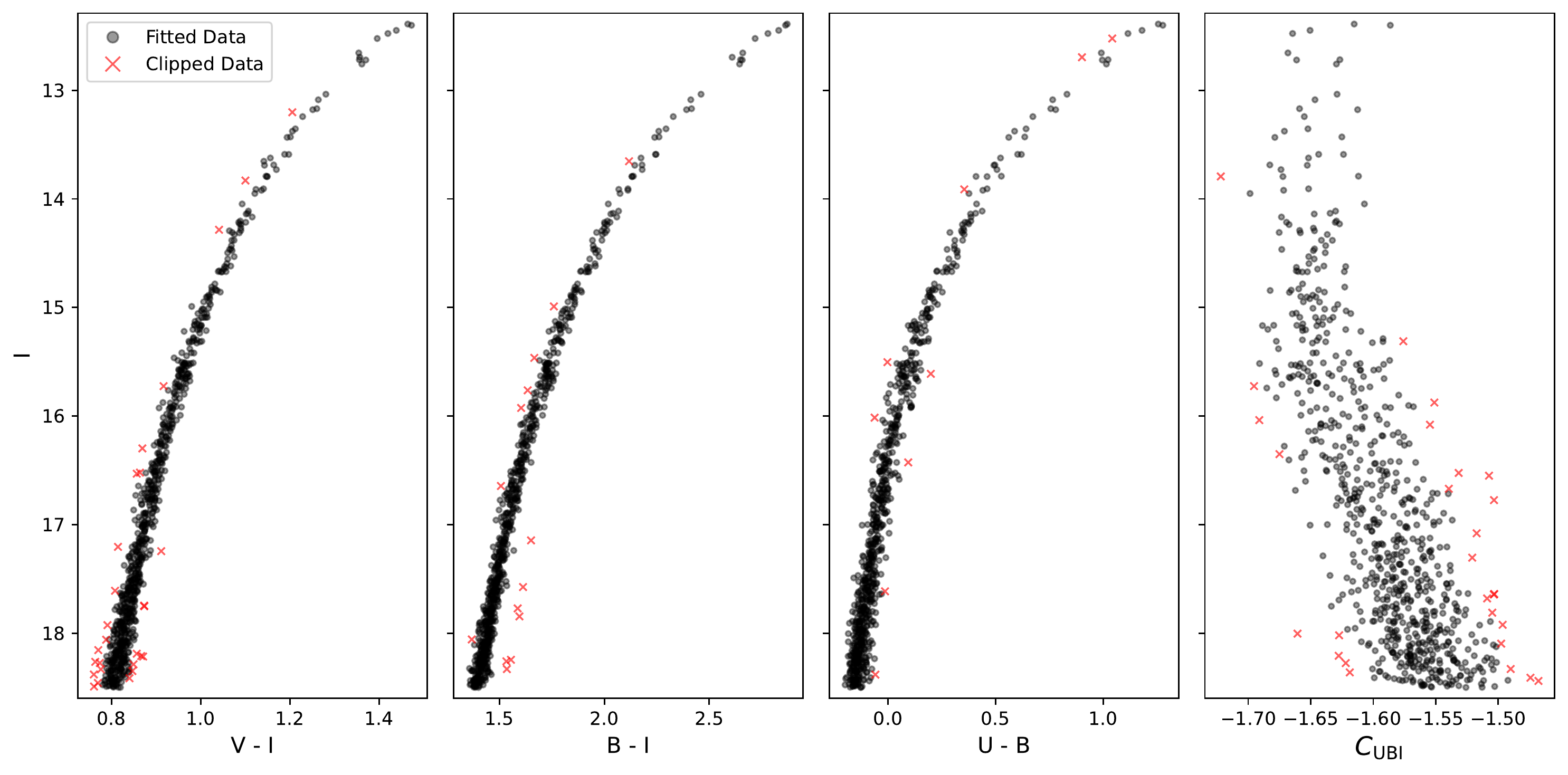}
    \caption{$N-\sigma$ clipping through various colour-index combinations for the ground-based photometry of NGC 5024. Outliers are shown in red, while stars that closely fit the polynomial applied to each distribution are shown in black. \textit{Right panel}: The same method was used on the $C_{\rm{UBI}}$ distribution.}
    \label{fig:nsigmaclips}
\end{figure*}

The purpose of the photometric cleaning process was to remove non-members and non-RGB stars so that the resulting distribution of RGB stars could be separated into multiple populations. We identified the Horizontal Branch (HB) and AGB stars in CMDs created from both the HST and ground-based photometry, as well as red and blue outlier stars that stray too far from the RGB. These stars were then manually removed from both sides of the RGB, allowing us to easily approximate and fit polynomials to the location of the RGB in the cluster CMD.

We applied a polynomial fit to the RGB in colour-colour and colour-magnitude diagrams using the Astropy LinearLSQFitter \citep{astropy}, so that outliers could be removed using an $N-\sigma$ clipping method. The colour-colour combination of $U-V$ vs $B-I$ ground-based bands shown in Figure \ref{fig:UVBI} was used for the polynomial fit, where the median ($m_{\mbox{fit}}$) was required (as opposed to the mean) as outliers surrounding the RGB stars can heavily affect mean values. Depending on the contamination of non-members and AGB stars in each cluster, the number of standard deviations to be cut from the median was adjusted within the range $2 \leq N \leq 3$. Highly contaminated clusters required a closer cut and therefore a smaller value of $N$. Non-members were identified according to $(U-V)_{\rm{obs}} - (U-V)_{\rm{fit}} > m_{\mbox{fit}} \pm (N \sigma_{\rm{(U-V)}})$, meaning all stars with a colour difference greater than $N$ standard deviations from the median of the polynomial fit were clipped. The process was iterated a maximum of three times. We also used this process for the HST photometry by using the closest equivalent colour-colour combination in the available HST bands.

We then applied the same 1D polynomial fitting and $N-\sigma$ clipping method to the following colour-index combinations in the ground-based photometry: ($V-I$), ($B-I$) and ($U-B$), and the HST photometry: ($F606W - F814W$), ($F438W - F814W$), ($F336W - F438W$) and ($F336W - F814W$). Finally, a special photometric index $C_{\rm{UBI}}$ was used, which separates stars based on their chemical properties, namely N and He abundances. $C_{\rm{UBI}}$ was first introduced by \cite{2013monelli} for ground-based photometry using Johnson filters with a focus on the RGB. It can also be adapted to the HST filters, as demonstrated by \cite{2013milonecubi}. For each star in the ground-based photometry: $C_{\rm{UBI}} = (U - B) - (B - I)$, while $C_{\rm{UBI}} = (F336W - F438W) - (F438W - F814W)$ in the HST photometry. We applied the same $N-\sigma$ clipping method on the resulting $C_{\rm{UBI}}$ distributions. The full sequence of polynomial fitting with $N-\sigma$ clipping is illustrated in Figure \ref{fig:nsigmaclips}, where red outliers were removed for each colour-index combination before finally removing outliers from the $C_{\rm{UBI}}$ distribution.

\subsection{Photometric Completeness Correction}
\label{sec:phot_incompleteness}

\begin{figure}
    \centering
    \includegraphics[width=\columnwidth]{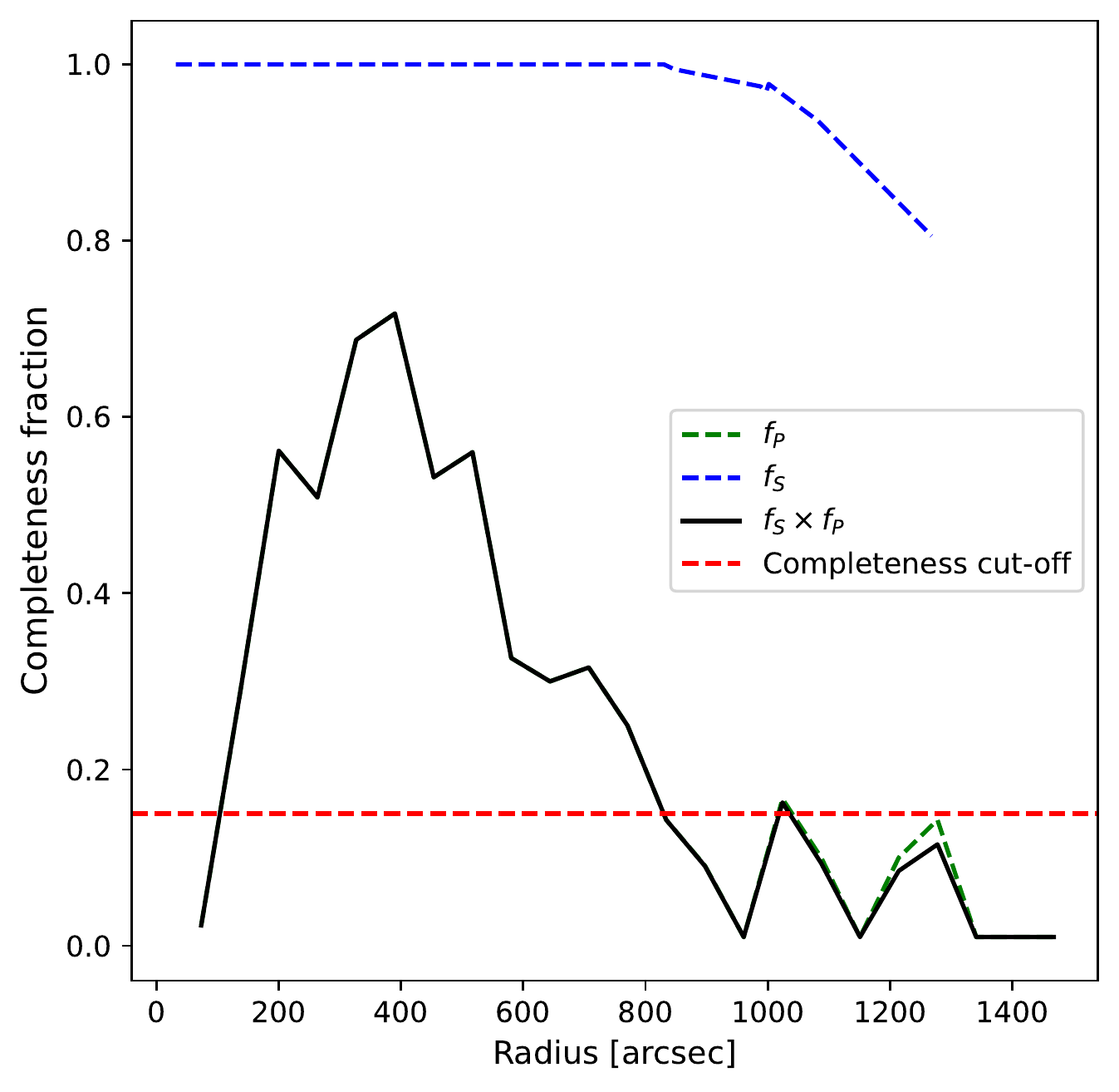}
    \caption{The individual spatial (blue) and photometric (green) completeness fractions, as well as the product of both completeness fractions ($f_S f_P$ in black) as a function of radius for the ground-based stars in NGC 5024. The dotted red line indicates the cut-off at $15\%$, which is the minimum accepted completeness fraction.}
    \label{fig:Nphot_comp}
\end{figure}
While the spatial completeness analysis of Section \ref{sec:spatial} compensates for cluster regions without observed stars caused by the limitations of the field, the photometric completeness compensates for a lack of stars due to poor or missing photometry. The aim is to assign a weighting to the surviving stars, such that they account for the fraction of stars that are lost during photometric cleaning. We assumed that both the HST and ground-based catalogues were complete at the magnitudes of the RGB, as \cite{2008Anderson} derives the completeness for the HST data as 100\% for stars brighter than the SBG for most clusters and \cite{stetsonphotometry} reports the ground-based data is complete across all radii for stars between $V = 19$ and $V = 12$. To determine the photometric completeness factor ($0 \leq f_P \leq 1.0$), we compared the number of RGB stars before and after the photometric cleaning processes. We divided the original spatial distribution of RGB stars radially into annuli and the number of stars before ($N_1$) vs. the number of stars after ($N_2$) determined the photometric completeness factor for stars in each annulus: $f_P = N_2 / N_1$. As we expect that the original HST and ground-based catalogues contain the vast majority of stars, this completeness factor accounts for the stars we remove in our cleaning, not stars missed by the catalogues.

The combined completeness fraction for each RGB star in both the HST and ground-based photometry was calculated as the product of the spatial and photometric completeness $f_T = f_S f_P$, which can be seen as a function of radius in Figure \ref{fig:Nphot_comp} for only the ground-based photometry of NGC 5024 as an example. The dense cluster center suffers a drop in completenesses due to the blending of stars in the ground-based catalogue, which were removed mainly through the quality cuts of Section \ref{sec:quality_indicators}. Additionally, the outer regions $R > 800 ''$ begin to drop in completenesses mainly due to the photometric cleaning of Section \ref{sec:photclean}. We stopped at the radius at which the combined completeness fraction dropped below $f_T < 0.15$ for the ground-based and HST photometry.

\subsection{Number Density Completeness}
\label{sec:magnitude_completeness}
\begin{figure}
    \centering
    \includegraphics[width=\columnwidth]{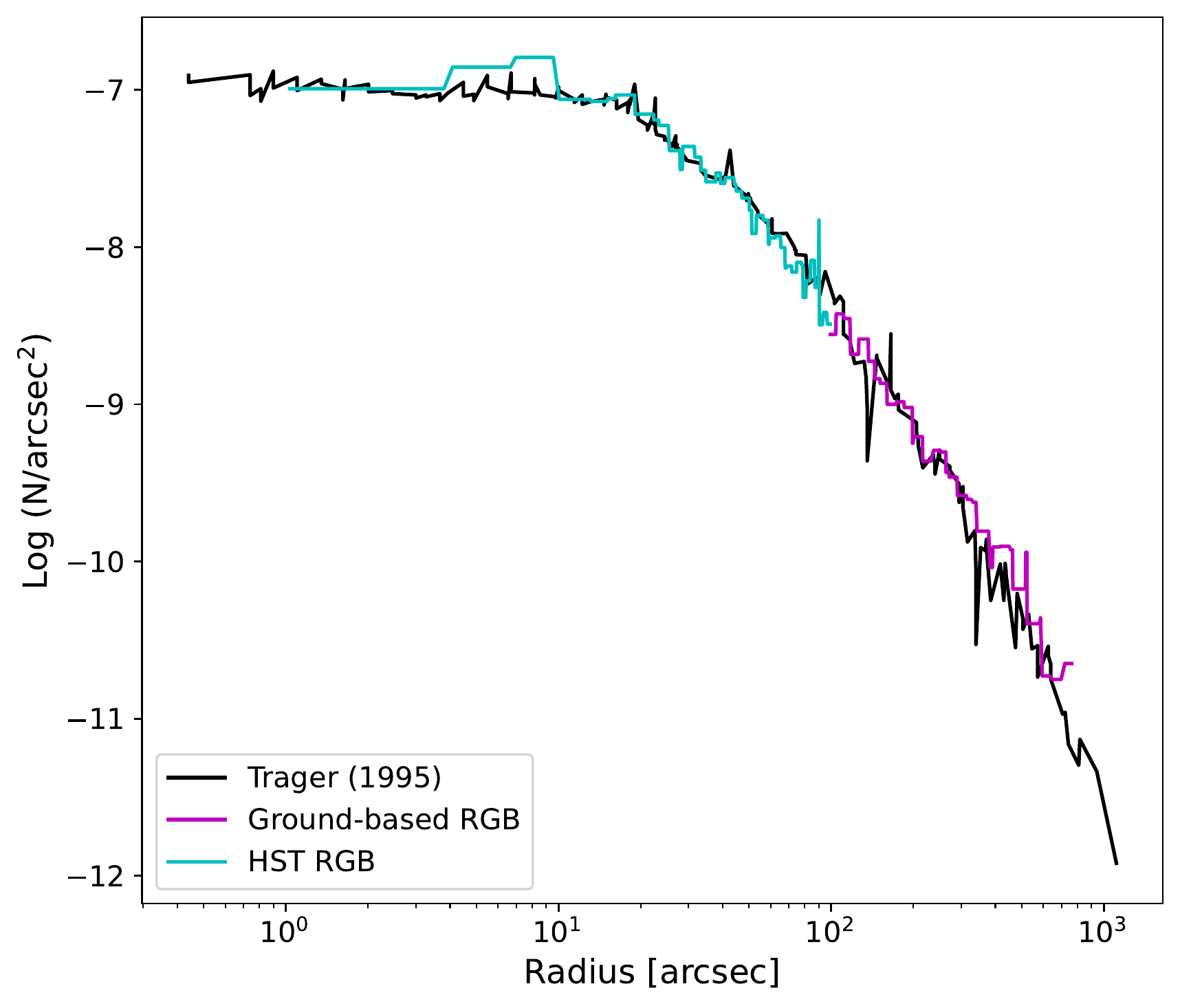}
    \caption{A comparison of the number density profiles as a function of radius for NGC 5024. The surface density profile from \citet{1995Trager} is shown in black. The HST cleaned and weighted RGB stars (cyan) transition into the ground-based cleaned and weighted RGB stars (magenta) at approximately $100''$ and matches well against the \citet{1995Trager}.}
    \label{fig:HSTtrager}
\end{figure}

In order to check the validity of our completeness corrections, we calculated the surface density based on completeness corrected stellar number counts and compared this against the surface brightness profiles of \cite{1995Trager}. The number density profile of the cleaned RGB stars in our sample was weighted by the spatial and photometric completenesses $f_T$. After correction for the combined completenesses, we applied the same shift factor to both the HST and ground-based data to convert between number density and surface density. The \cite{1995Trager} data was used as a reference profile and is shown in black in Figure \ref{fig:HSTtrager}. We then compared the number density profiles of the HST (cyan) and ground-based data (magenta) to the reference profile for each cluster in order to confirm the viability of the total incompleteness factors as a weighting to compensate for missing photometry. We found a good match between the HST and ground-based photometry and the \cite{1995Trager} profile for all 28 GCs in our sample.

\section{Identification of Multiple Populations}
\label{sec:popsplit}
We now move on to separate the multiple stellar populations using both the $C_{\rm{UBI}}$ distribution method (Section \ref{sec:GMM}) and the chromosome map method (Section \ref{sec:chromosome}), before finally analysing their radial distributions (Section \ref{sec:Aplus}).

\subsection{Gaussian Mixture Models applied to $C_{\rm{UBI}}$ Distributions}
\label{sec:GMM}

\begin{figure}
    \centering
    \includegraphics[width=\columnwidth]{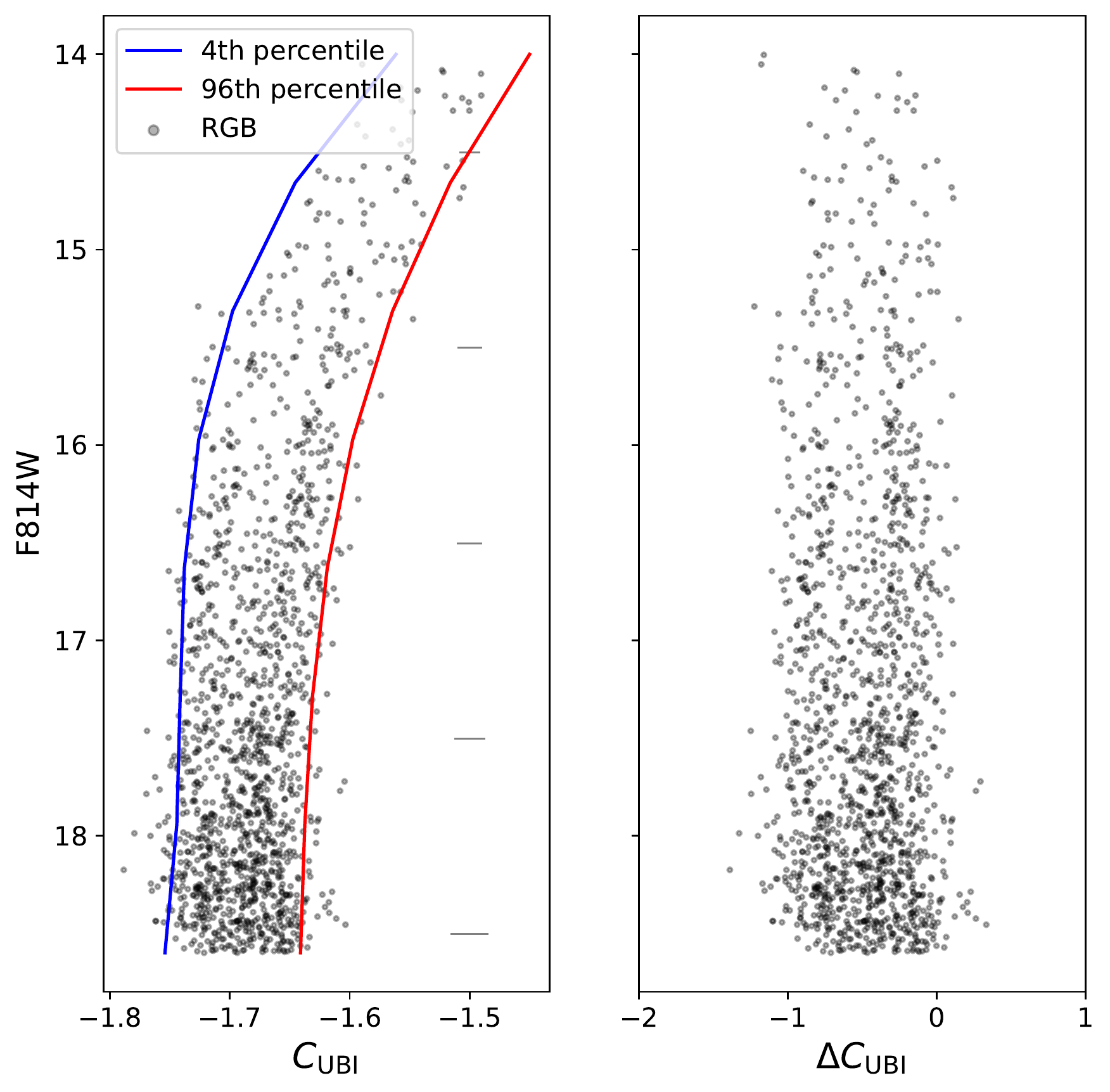}
    \caption{\textit{Left panel}: The $C_{\rm{UBI}}$ distribution of NGC 5024 stars in black using HST photometry, with the $4^{th}$ percentile ridgeline in blue and the $96^{th}$ percentile ridgeline in red. Grey horizontal lines indicate the photometric error in the $C_{\rm{UBI}}$ distribution at different magnitudes. \textit{Right panel}: The resulting distribution $\Delta C_{\rm{UBI}}$ of the same stars after normalisation as described by Equation \ref{eq:ridgeline}.}
    \label{fig:ridgelines}
\end{figure}

\begin{figure*}
    \centering
    \includegraphics[width=\textwidth]{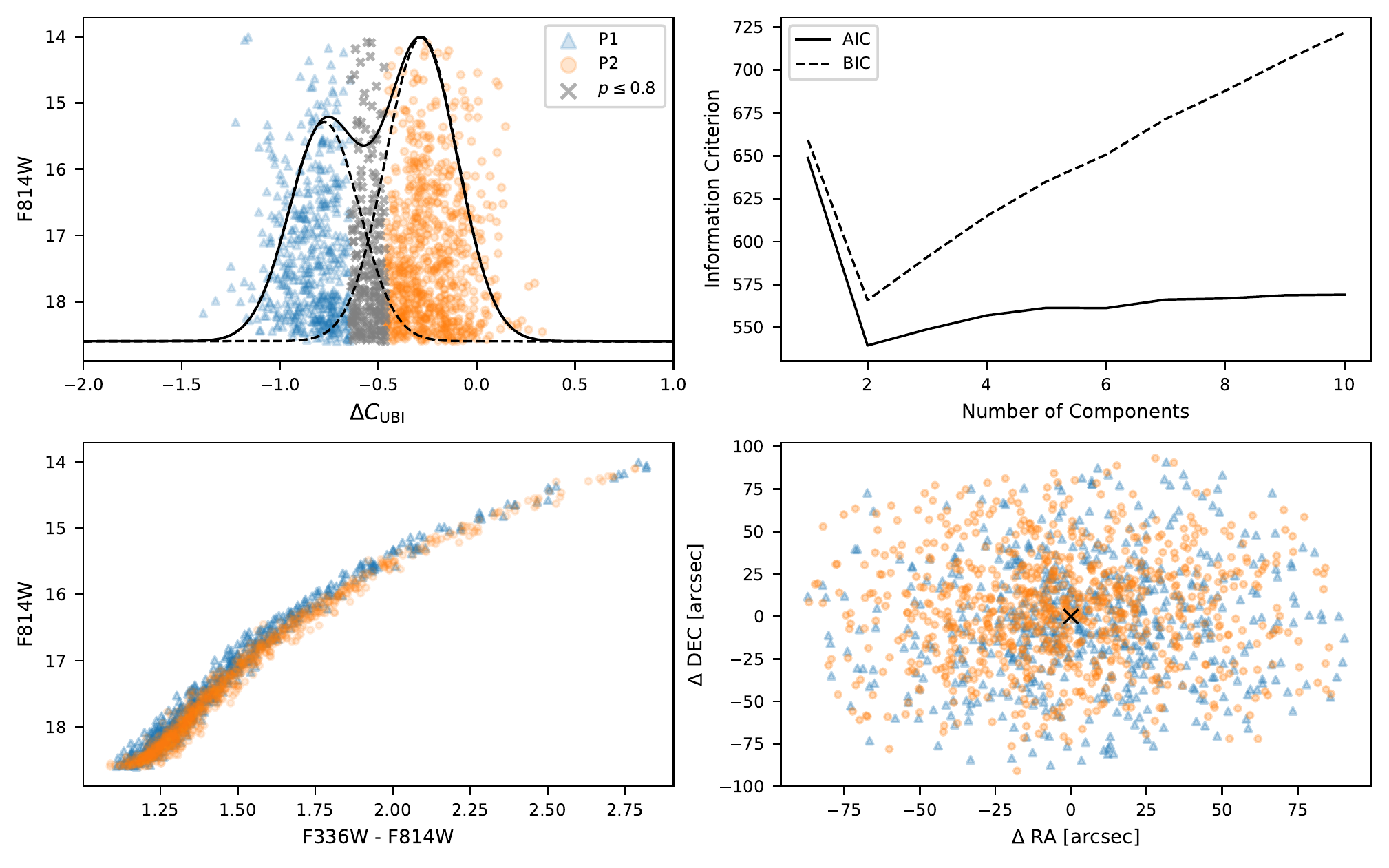}
    \caption{Population separation of NGC 5024 using HST photometry. \textit{Top left}: The best-fit GMM (solid black line) with the corresponding individual Gaussians (dashed), together with the $\Delta C_{\rm{UBI}}$ distribution of stars separated into their respective P1 and P2 populations. In grey we show stars with ambiguous classification, i.e. membership probability to either population of $p \leq 0.8$. \textit{Top right}: The AIC and BIC both show a minimum at $n=2$, indicating a clear identification of two populations. \textit{Bottom left}: The CMD of the two populations from the MS turn-off to the tip of the RGB. \textit{Bottom right}: The spatial distribution of the two populations showing isotropic behaviour.}
    \label{fig:4panelplot}
\end{figure*}

The multiple stellar populations of each cluster were identified using the photometric index $C_{\rm{UBI}}$ described in Section \ref{sec:photclean}. The general method for categorizing stars into multiple populations throughout this paper involved applying Gaussians to the $\Delta C_{\rm{UBI}}$ distribution of stars, which is a normalised version of the $C_{\rm{UBI}}$ distribution as shown in Figure \ref{fig:ridgelines}. To normalise the distribution, the $4^{th}$ and $96^{th}$ percentiles of the combined $C_{\rm{UBI}}$ values for all stars were determined and fitted with a 1D polynomial, as per the method detailed in \cite{2017milone}. We used Equation \ref{eq:ridgeline} to calculate the normalised distribution $\Delta C_{\rm{UBI}}$ from the distributions of $C_{\rm{UBI}}$ in both the HST and ground-based photometry.
\begin{equation}
    \Delta C_{\rm{UBI}} = \frac{C_{\rm{UBI}} - X_{blue}[I]}{X_{red}[I] - X_{blue}[I]} - 1
    \label{eq:ridgeline}
\end{equation}
The red $(X_{red})$ and blue $(X_{blue})$ fiducial ridgelines in the left panel of Figure \ref{fig:ridgelines} were created at equally sized increments of $F814W$ and $I$ magnitude bins for the HST and ground-based photometry, respectively. An example of the resulting $\Delta C_{\rm{UBI}}$ distribution is shown in the right panel of Figure \ref{fig:ridgelines}. We note that for all clusters in our sample, the photometric error in the $C_{\rm{UBI}}$ distribution is much smaller than the colour spread in $C_{\rm{UBI}}$ due to the presence of multiple stellar populations. Due to this, we are confident that the separation between multiple populations in the $C_{\rm{UBI}}$ distribution is not influenced by photometric errors in the bands.\\

With this normalised distribution of stars, Gaussian Mixture Models (GMMs) from the scikit-learn package \citep{scikit-learn} were applied in order to find the most probable distribution of the mutliple populations. The method uses an expectation-maximization approach in order to determine the best mixture of one or more Gaussians to fit the $\Delta C_{\rm{UBI}}$ distribution. Both the Akaike information criterion (AIC) and Bayesian information criterion (BIC) were used to determine the most probable number of populations when provided with the $\Delta C_{\rm{UBI}}$ distribution of a cluster. The minima of both the AIC - which estimates the relative quality of the statistical models based on in-sample prediction error, and the BIC - which selects the most probable model based on likelihood functions, indicated the most probable number of populations within a cluster from a range of $1 \leq n \leq 10$ different components. For most clusters the AIC and BIC found $n=2$ components. The top right panel of Figure \ref{fig:4panelplot} shows the range of possible components when applying GMMs to NGC 5024, with both AIC and BIC providing minima at $n=2$. Clusters with minima at $n=1$ were discarded.

From the most probable GMM samples, the final separation of the populations was created in terms of two or more Gaussians encompassing the full sample of stars. The top left panel of Figure \ref{fig:4panelplot} shows the combination of two Gaussians on the $\Delta C_{\rm{UBI}}$ distribution of stars. Each star was assigned to a population based on the probability that it belonged to a particular Gaussian. This membership probability was also used to divide the multiple populations for clusters with three populations, as discussed further in Section \ref{sec:chromosome}. We required stars to have membership probability $p \geq 0.8$ between the P1 and P2 populations. This resulted in a small gap between each of the Gaussians, shown as gray points in Figure \ref{fig:4panelplot}, ensuring that the stars belong to the population they were assigned to with high confidence. We experimented with this threshold using $0.5 \leq p \leq 1.0$ in increments of 0.05 and found the overall results and conclusions of this work were not affected by the exact value of the threshold. Similarly, we tested the effect of changing the limit of the primordial and enriched classifications for clusters with interesting radial distributions\footnote{NGC 3201, NGC 6101 and NGC 7078 -- see Section \ref{sec:P1conc}}. Briefly, we randomly sampled arbitrary limits in the $\Delta C_{\rm{UBI}}$ colour (i.e. the point where the Gaussians overlap) and classified stars left of the limit as primordial and stars to the right as enriched. The limit was drawn from a uniform distribution covering the inner $2\sigma$ of the $\Delta C_{\rm{UBI}}$ colour to avoid a cut too close to either colour end. We did this to prevent having almost all stars classified into one population with only a few left to be classified in another. For the purpose of these tests, we continued the remainder of the analysis with these arbitrary classifications in order to statistically determine the significance of our resulting radial profiles. We sampled the arbitrary limits 200 times per cluster and each time we sampled anywhere from 90 to 100\% of the stars on either side of the limit to also observe the effect of randomly removing individual stars from each population.

\subsection{Chromosome Maps}
\label{sec:chromosome}

\begin{figure}
    \centering
    \includegraphics[width=\columnwidth]{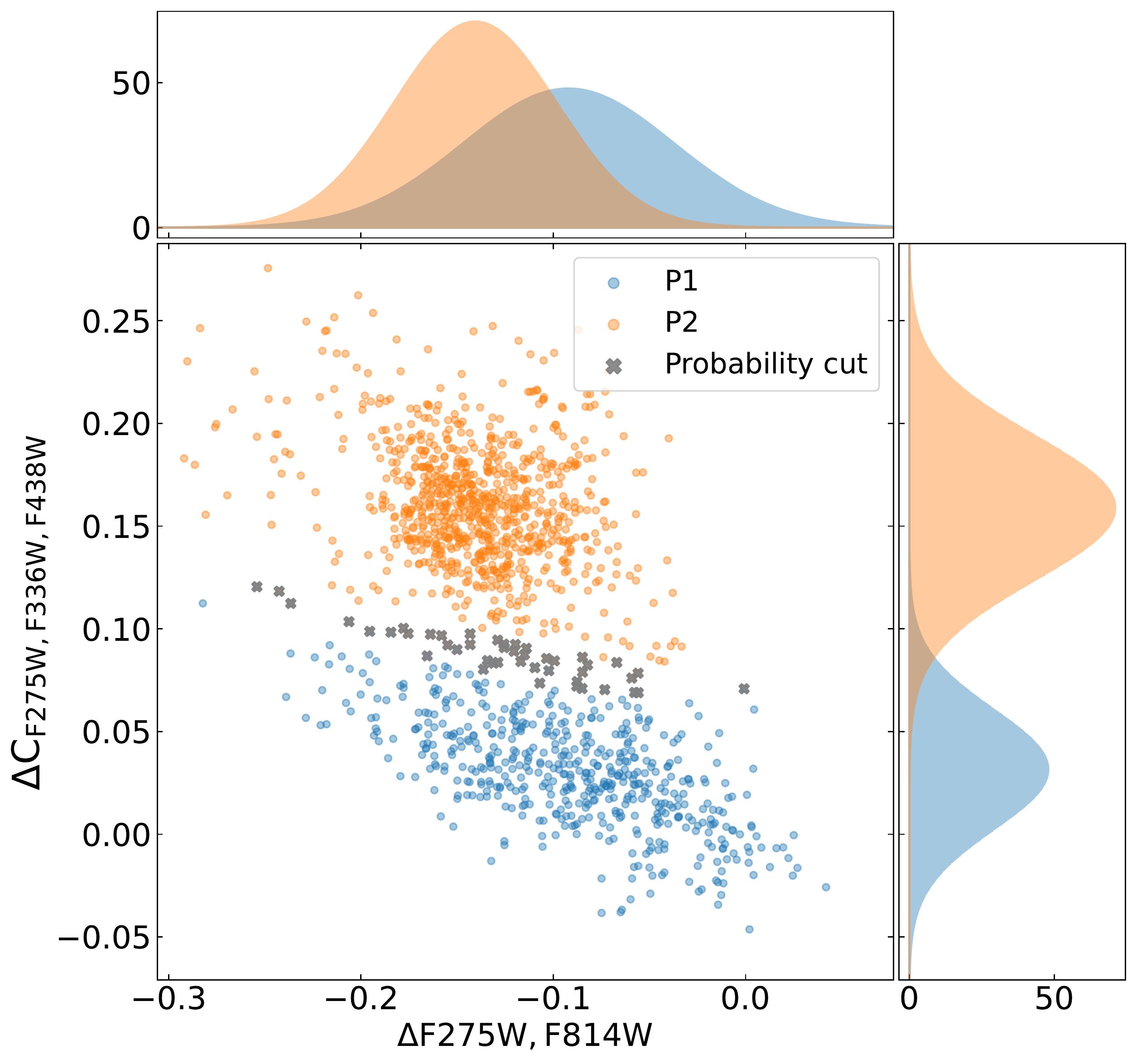}
    \caption{Chromosome map using the HST photometry for NGC 5024, with Gaussian Mixture Models (GMMs) applied in two dimensions. The lower left plot shows the chromosome map with populations P1 (blue) and P2 (orange) as defined by the two Gaussians in the top and right panels. In grey are stars which lie in-between the two populations, with membership probabilites $p\leq0.8$ for either population.}
    \label{fig:chromosome}
\end{figure}

In addition to the $C_{\rm{UBI}}$ colour distribution classification, for the HST photometry it is also possible to separate the populations using chromosome maps. Introduced by \cite{2017milone}, a chromosome map is a colour-colour plot that has been normalised in a way which allows efficient separation of sub-populations of different abundances. It uses the RGB width in a $F275W - F814W$ vs. $F814W$ CMD, along with the RGB width of the pseudo-colour combination $C_{F275W,F336W,F438W}$ vs. $F814W$. Following the method in \cite{2017milone}, we defined a dividing line between populations in the $\Delta_{ F275W,F814W}$ vs. $\Delta C_{F275W,F336W,F438W}$ distribution. We found that clusters such as NGC 2808 contained several distinct populations which can be split using a chromosome map. In these instances, the multiple populations tend to be easier to distinguish using a chromosome map, as they can become somewhat blended together when using the $\Delta C_{\rm{UBI}}$ distribution alone. Therefore, by creating chromosome maps and then using the GMM method in two dimensions, as shown in Figure \ref{fig:chromosome}, we were able to directly compare the populations separated using a $\Delta C_{\rm{UBI}}$ plot, against the populations separated by a chromosome map. The aim was to implement the same membership probability defined in Section \ref{sec:GMM} of $p \geq 0.8$ to cut out the ambiguous stars, shown in grey in Figure \ref{fig:chromosome}, before checking how the remaining stars were assigned to populations according to the two methods.

\begin{figure}
    \centering
    \includegraphics[width=\columnwidth]{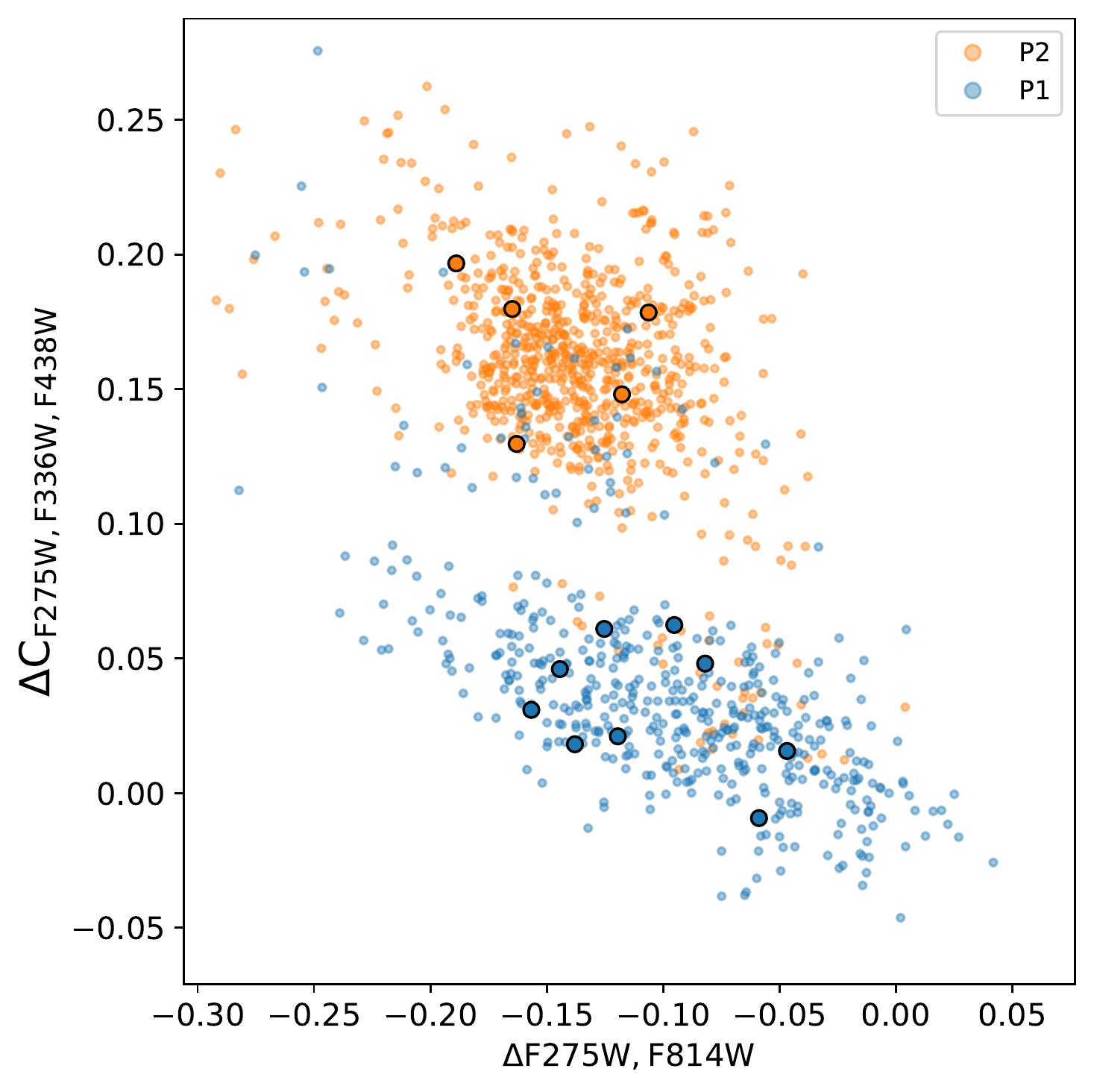}
    \caption{Chromosome map using the HST photometry for NGC 5024, as shown in Figure \ref{fig:chromosome}; however, we now use the $\Delta C_{\rm{UBI}}$ separation to assign the stars into P1 (blue) and P2 (orange) populations. Stars with membership probabilities $p\leq0.8$ are also removed. There is still a very good separation, as also shown in Figure \ref{fig:chromosome}, so we can see `contaminant' stars by eye as blue points located in the orange clump and vice versa. Bold circles indicate stars that overlap in both the HST and ground-based photometry, colour-coded to show the agreement between their independent classifications in each photometric catalogue.}
    \label{fig:chrome_gmm_map}
\end{figure}

The HST photometry includes the UV filter F275W which has no ground-based equivalent. We therefore relied on the $C_{\rm{UBI}}$ distribution of the HST and ground-based photometry for a consistent analysis. The HST F275W photometry was only used to confirm whether the $C_{\rm{UBI}}$ classification was consistent with the chromosome map method. To do this, the RGB stars of the $C_{\rm{UBI}}$ distribution were separated into multiple populations with both methods. In Figure \ref{fig:chrome_gmm_map} we show the chromosome map of NGC 5024, where we colour code the stars classified as P1 and P2 with the $\Delta C_{\rm{UBI}}$ distribution method in orange and blue, respectively. This figure shows that for the majority of the stars, the classification of different populations using $\Delta C_{\rm{UBI}}$ was consistent with the classification based on the chromosome map.

In all clusters, there was a small percentage of stars where the P1/P2 classification obtained using the chromosome map and $\Delta C_{\rm{UBI}}$ disagree. We see that there are $\Delta C_{\rm{UBI}}$ P1 stars in Figure \ref{fig:chrome_gmm_map} (blue) that inhabit the region in which the bulk of the P2 stars (orange) are located, and vice versa. We found the average fraction of stars that were classified differently by each method was $\sim10\%$ for the 28 GCs in our final sample, with a minimum of 4\% and a maximum of 20\% after the probability cut. Clusters with high contamination percentages had heavily blended populations in the chromosome map, meaning the distribution of stars followed a more continuous distribution as opposed to distinct clumps. This caused difficulties in accurately determining the classification of populations in one or both separation methods and therefore these clusters were excluded from our analysis. To further check the consistency of the population classification, we used overlapping stars that were covered by both (ground based and HST) photometric catalogues and had been independently classified into the different sub-populations using each data set. We found consistent classifications of populations for stars common to both data sets, as demonstrated with large bold blue (P1) and orange (P2) points in Figure \ref{fig:chrome_gmm_map}.\\

After ensuring consistent results between the different classification methods/catalogues, we combined the HST and ground-based photometry by removing stars in the ground-based data which overlap with the HST field. By doing this, we ensure the ground-based data begins at the same radius where the HST data ends, ensuring there are no gaps between the fields. We then use this combined data set to study the behaviour of MPs across the full extent of these clusters.

\subsection{Radial Distributions of Different Populations}
\label{sec:Aplus}

\begin{figure*}
    \centering
    \includegraphics[width=\textwidth]{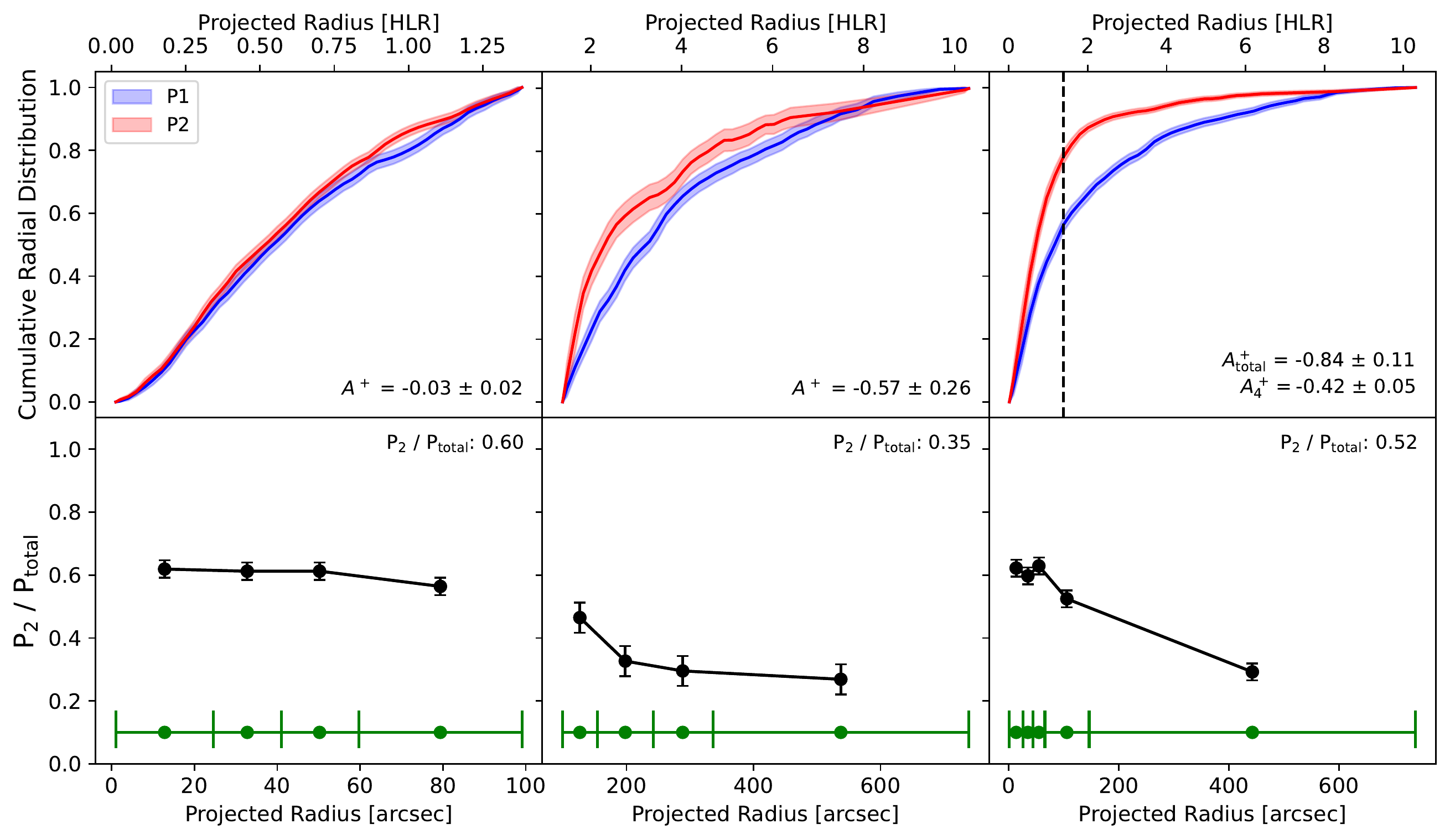}
    \caption{Cumulative radial distributions of different populations and enriched star fractions in NGC 5024 for the full radial range of the cluster. The scale shown for reference in the top panels is the distance from the centre of the cluster in units of projected half light radius [HLR]. \textit{Upper left:} The weighted, normalised, cumulative radial distribution of P1 (blue) and P2 (red) stars in the HST photometry within $r < 100''$ of the cluster center. \textit{Upper middle:} The ground-based photometry from $100'' < r < 740''$, analysed in the same way as the HST data in the upper left panel. \textit{Upper right:} The $A^+$ parameter for the combined data set, covering $0 < r < 740''$. The radius at which the HST photometry meets the ground-based photometry is shown by a black, dashed line. We quote both $A^+_{4}$ for the calculated $A^+$ value at a radial limit of 4.27$R_{\rm{hlp}}$ and $A^+_{\rm{total}}$ for the full radial range. \textit{Lower left:} The fraction of P2 stars as a function of radius for the HST photometry (black). Each bin has an equal number of stars, with the radial range of the bins illustrated at the bottom of the plot (green). \textit{Lower middle:} The ground-based photometry analysed in the same way as the HST data in the lower left panel. \textit{Lower right:} The P2 fraction as a function of radius for the combined data set. The total $\rm{P_2/P_{total}}$ fractions are indicated in each panel for each corresponding radius range.}
    \label{fig:Aplus}
\end{figure*}

A useful tool in understanding the behaviour of MPs as a function of radius is calculating the cumulative radial distribution of the stars in each population. If one population is more centrally concentrated within the cluster, we see a comparatively steeper slope in its cumulative radial distribution than we do for the other population. However, if the populations are homogeneously mixed throughout the cluster, we see similar slopes for both distributions. The $A^+$ parameter introduced by \cite{2016Aplus} is a way to quantify differing radial profiles, as it is an integration of the `area' between the two distributions.
The cumulative radial distributions in this work provide a spatially complete view of each cluster by combining the innermost region using HST photometry with the outer region using ground-based photometry. To calculate the cumulative radial distributions we used the method introduced and detailed by \cite{2016Aplus} and \cite{2019dalessandro}. The $A^+$ parameter considers the area between the cumulative radial distributions of \textit{two} populations, so for clusters exhibiting \textit{three} distinct stellar populations such as NGC 1851, NGC 2808, NGC 6101 and NGC 7078, we combined the P2 and P3 stars into a single `enriched' population, referred to as P2 for simplicity. This classification follows the logic of \citet{2017milone}, in which the primordial stars (P1) are identified as the group of stars aligning with $\Delta C_{F275W, F336W, F438W} = \Delta F275W, F814W = 0$ in a chromosome map, while P2 stars are any stellar populations located above the primordial stars.

We calculated a modified version of the $A^+$ parameter using Equation \ref{eq:aplus} in order to characterize the \textit{weighted} cumulative radial distributions of stars in each population using the total completeness fractions $f_T$ calculated in Section \ref{sec:phot_incompleteness}.
\begin{equation}
    \label{eq:aplus}
    A^+ (R) = \int^R_{R_{min}} \left( \phi_{P1} (R') - \phi_{P2} (R') \right) dR'
\end{equation}
Here, $\phi$ is the normalised, cumulative sum of the weights, $w = 1/f_T$, of the stars in either the P1 or P2 population. Our $A^+$ parameter indicates whether a cluster has a P1 concentration in the center ($A^+ > 0$), a P2 concentration in the center ($A^+ < 0$), or a homogeneous mix of populations ($A^+ \sim 0$) throughout the cluster. The uncertainty in $A^+$ was determined via bootstrapping. Briefly, the P1 and P2 stars of each cluster were sampled randomly for a total of 500 iterations using a sample size of 1000, with an $A^+$ value calculated each time. The final uncertainty for each cluster was calculated from the standard deviation of the 500 iterations.

Figure \ref{fig:Aplus} shows the weighted and normalised cumulative radial distributions of the two stellar populations found in NGC 5024 along the top panels, with the bottom panels showing the corresponding number ratio of enriched to total stars ($\rm{P_2 / P_{total}}$) as a function of radius. NGC 5024 is an example of why the full extent of the cluster should be analysed when considering the radial distributions of populations within a cluster. The left panels show the behaviour of the cluster for only the HST field (1293 stars). We already see by eye that both cumulative profiles are almost identical, which is also supported numerically by the parameter $A^+ = -0.03 \pm 0.02$. The cumulative radial distribution of the HST photometry alone would suggest that the populations of this cluster are fully mixed and spatially indistinguishable. However, the middle panels show the result of this same analysis on the ground-based photometry (438 stars). Here P2 is more centrally concentrat‹ed ($A^+ = -0.57 \pm 0.26$), with the outer regions dominated by P1 stars. Finally, in the right panel, the full extent of the cluster is analysed by combining both the HST and ground-based stars, producing a value of $A^+ = -0.84 \pm 0.11$ and supporting the result that P2 is centrally concentrated. This information is lost when only observing the cluster center and using the resulting $A^+$ parameter to describe the behaviour of the cluster as a whole. It is especially important to consider the outer regions of clusters, since dynamical mixing of the populations will affect the center of the cluster within shorter timescales than it does for the outer stars \citep{2019dalessandro}. To show the consistency of behaviour between the two photometric data sets, we plot the enriched star fraction $\rm{P_2 / P_{total}}$ as a function of radius in the lower panels of Figure \ref{fig:Aplus}. Here, the inner region also shows a mostly constant P2 concentration and the outer region shows a strong decline in P2 stars, supporting the result of the cumulative radial distributions while also showing agreement in the transition region between data sets.

\section{Results}
\label{sec:results}

\begin{figure*}
    \centering
    \includegraphics[width=\textwidth]{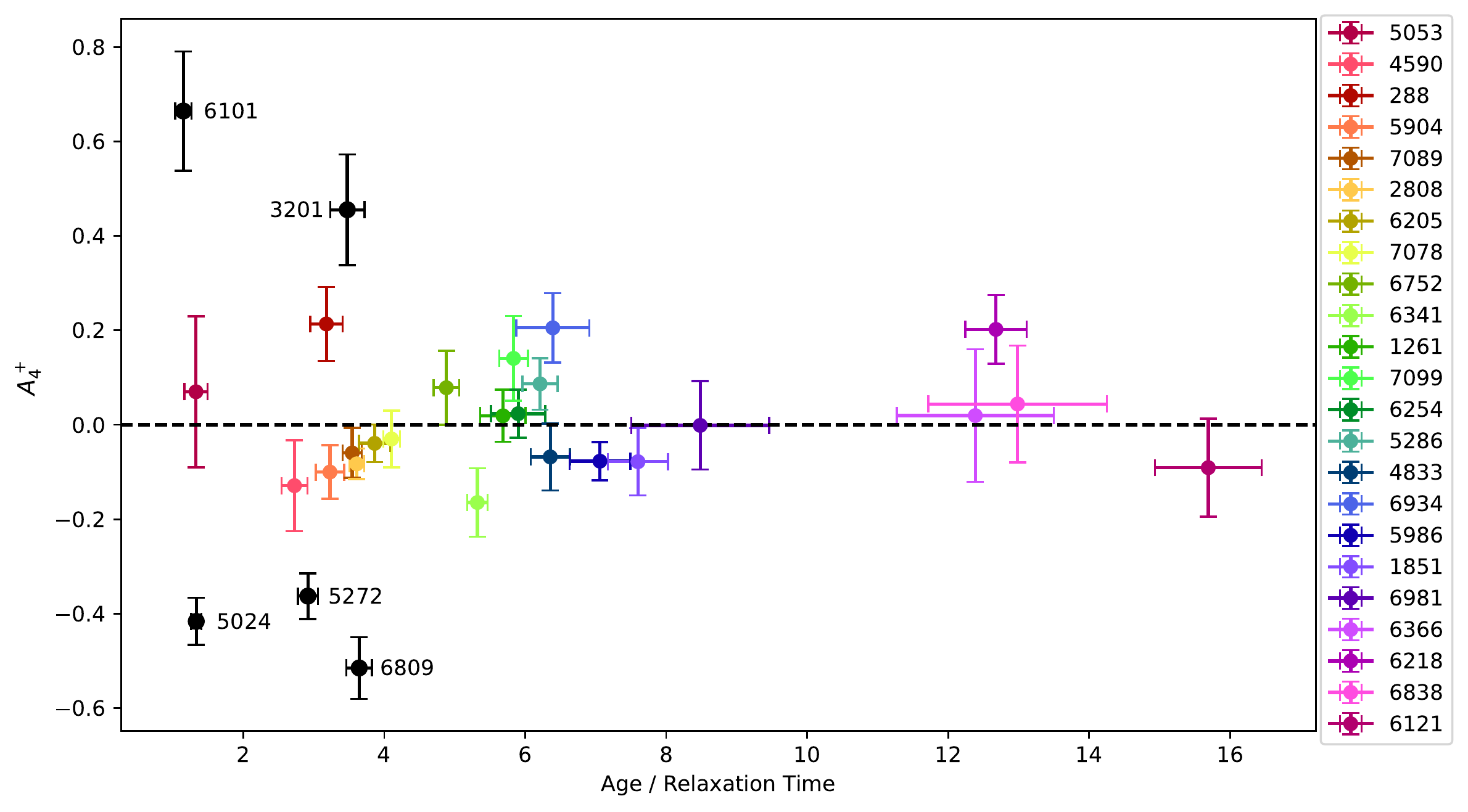}
    \caption{The total cumulative radial distributions in terms of the $A^+_{4}$ parameter for the 28 Galactic GCs as a function of their dynamical age. All clusters are limited to a radius equivalent to 4.27$R_{\rm{hlp}}$ for direct comparison. Using this radius limit, clusters with an $A^+$ value greater than 3-$\sigma$ significance from zero are displayed as labelled black points. An $A^+$ value close to zero indicates the MPs are spatially mixed throughout the analysed spatial extent of the cluster. Significantly positive $A^+$ values indicate that the primordial (P1) population is more centrally concentrated, while negative values indicate the enriched (P2) population is more centrally concentrated. Refer to Section \ref{sec:Apluszero} for a special discussion on NGC 7078.}
    \label{fig:Aplustr}
\end{figure*}

\begin{figure*}
    \centering
    \includegraphics[width=\textwidth]{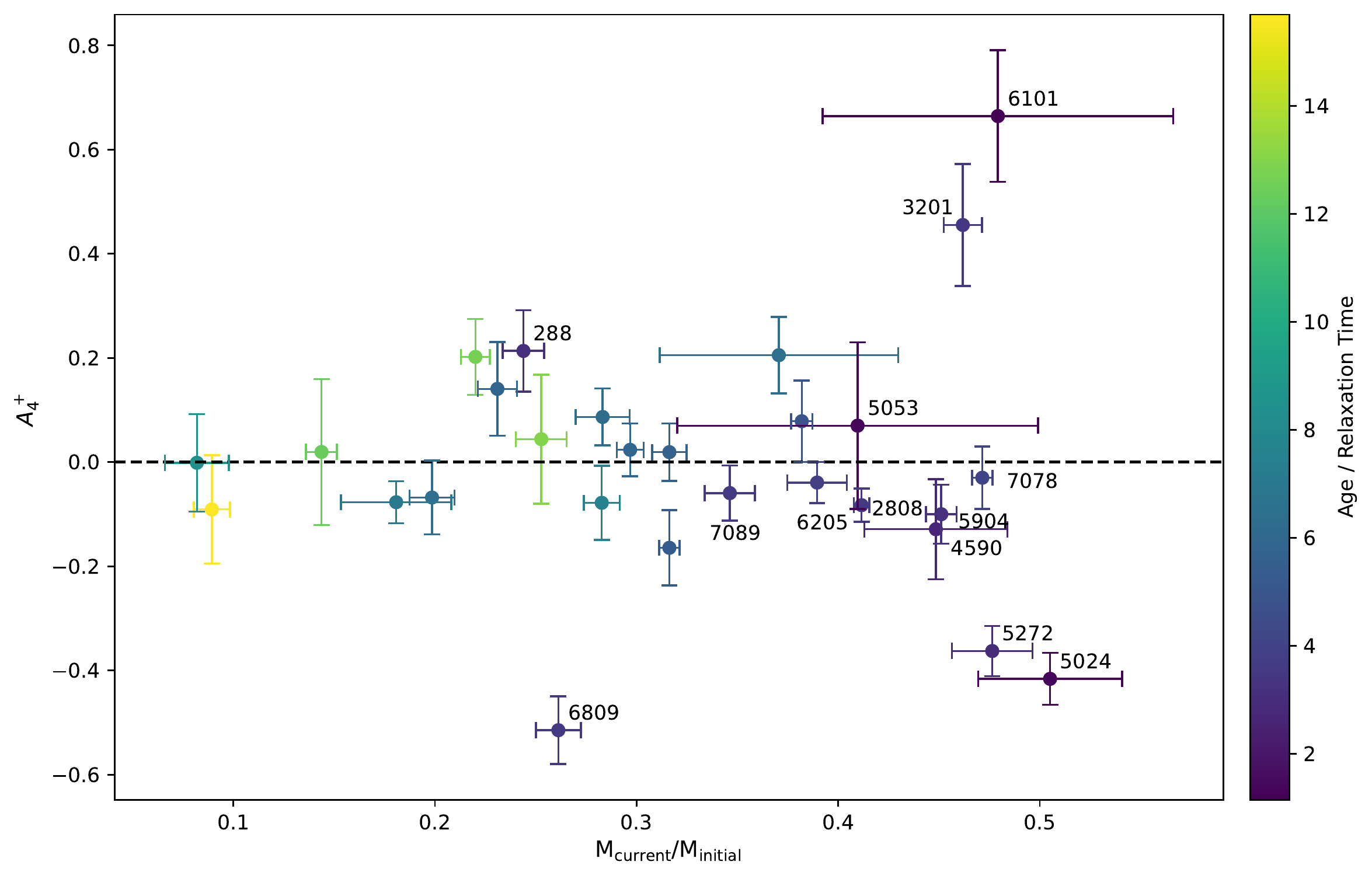}
    \caption{The total cumulative radial distributions in terms of the $A^+_{4}$ parameter (calculated at a maximum radius of 4.27$R_{\rm{hlp}}$) for the 28 Galactic GCs as a function of their mass loss ratio. Each cluster is also colour-coded by its dynamical age. Clusters categorised as `dynamically young' (age/relaxation time < 4.5) are displayed as labelled points. Refer to Section \ref{sec:Apluszero} for a special discussion on NGC 7078.}
    \label{fig:Aplusmassloss}
\end{figure*}

For the 28 Galactic GCs in our sample we now investigate the trends associated with the $A^+$ parameter and the enriched star fraction $\rm{P_2/P_{total}}$. In Section \ref{sec:Aplusresults} we explore the global trends using the cumulative radial distributions, in Section \ref{sec:pfracresults} we explore the global trends using the enriched star fractions $\rm{P_2/P_{total}}$, and finally in Section \ref{sec:notableclusters} we discuss individual notable clusters that have low dynamical ages.\\

Throughout this section we use cluster parameters provided by the Galactic Globular Cluster Database by \cite{2019baumgardthilker}, updated to the Gaia DR3 data as described by \cite{clusterdatabase} and \cite{2021baumgardt}. We take the initial cluster mass and current cluster mass values, the former being calculated from the current cluster masses and cluster orbits using Equation 3 from \cite{2003baumgardt}. The relaxation time ($T_{RH}$) of each cluster was also used, giving the time scale in which each cluster will become dynamically mixed, which was derived by \cite{2018baumgardthilker}. We define the dynamical age as the ratio of the age of a star cluster to its relaxation time and estimate the mass loss ratio ($\rm{M_{c} / M_{i}}$) as the ratio of the current ($\rm{M_{c}}$) and initial ($\rm{M_{i}}$) mass of the cluster. We also take the projected half-light radius ($R_{\rm{hlp}}$), half-mass radius and orbital parameter values for each cluster from this database. The cluster ages are taken from the work of \cite{2019Kruijssen}, while metallicity values are taken from \cite{2010harris}.

Previous work has found a clear correlation of the width of the RGB in clusters with MPs as a function of cluster metallicity $[\rm{Fe/H}]$, absolute visual magnitude $M_V$ and initial mass of the cluster \citep{2013monelli,2017milone}. Since we have combined two independent photometric catalogues to get an extended spatial view, it was important that we replicated the well-established trends observed by others who used the same catalogues. In particular, we followed the method set out by \cite{2013monelli} for the ground-based catalogue and determined the RGB widths ($\rm{W_{RGB}}$) in the same manner for the 28 Galactic GCs in our sample. We found a strong correlation between $\rm{W_{RGB}}$ and $[\rm{Fe/H}]$, with a Spearman correlation coefficient $r_s = 0.693$ and associated p-value $= 4\times10^{-5}$, as well as an anti-correlation between $\rm{W_{RGB}}$ and $M_V$, with $r_s = -0.331$ and a p-value $ = 0.08$. For the HST data, we followed the method of \cite{2017milone} and reproduced the correlation between $\rm{W_{F275W,F814W}}$ and $[\rm{Fe/H}]$ for clusters with $M_V > -7.3$, providing a Spearman correlation coefficient of $r_s = 0.704$ and a p-value $ = 4\times10^{-5}$. We also reproduced the trend between $\rm{W_{F275W,F814W}}$ and $M_V$, with $r_s = -0.104$ and p-value $ = 0.6$. We conclude that our data exhibits the same well-established trends as previous work.

\subsection{Global Trends using Cumulative Radial Distributions ($A^+$)}
\label{sec:Aplusresults}

We analysed large regions of the targets in our sample of 28 Galactic GCs and calculated the cumulative radial distribution parameters $A^+$. We then identified clusters in which the $A^+$ values indicated a high central concentration of either primordial or enriched stars at a significance larger than $3\sigma$. These significantly segregated clusters will be discussed in detail in Sections \ref{sec:P2conc} and \ref{sec:P1conc}. The maximum radii for the outermost stars in the ground-based fields differed greatly for each cluster, so in order to make the results in different clusters comparable to each other, we analysed the spatial distribution of stars only out to $4.27 R_{\rm{hlp}}$ in all clusters. We chose this limit as it was the minimum radius for our final sample of stars in NGC 6101, with most clusters extending beyond this radial limit. The only clusters that did not reach this limit were NGC 3201, NGC 5053, NGC 6121 and NGC 6838 where the maximum radii for the ground-based photometry were in the range of $2.5 R_{hlp}$ (NGC 6838) $< r_{max} < 4.0 R_{hlp}$ (NGC 3201). Limiting all clusters to this lower range would remove important information on the cluster properties in the outermost regions. Therefore, for these four clusters we assumed that the relative fraction of primordial and enriched stars is constant from the outermost radius covered by our photometry to $4.27 R_{\rm{hlp}}$. Since we extrapolate out to $4.27 R_{\rm{hlp}}$ by sampling real stars in the outer radial bins, we do not expect that this will add significant uncertainty to the $A^+$ parameters as we also propagate the uncertainties of these sampled stars.

Figure \ref{fig:Aplustr} shows the resulting $A^+_{4}$ parameters (calculated at a maximum radius of $4.27R_{\rm{hlp}}$) as a function of dynamical age. We found that dynamically old clusters all have $A^+ \sim 0$, in agreement with the idea that due to relaxation, populations become mixed. This also agrees with the findings of \cite{2019dalessandro}. 

In dynamically young clusters, we found a larger range of $A^+$ values. Surprisingly, we not only found centrally concentrated P2 populations ($A^+ < 0$, e.g. NGC 2808, NGC 5024, NGC 5272 and NGC 6809) consistent with the findings of \cite{2019dalessandro}, but also clusters with centrally concentrated P1 populations ($A^+ > 0$, e.g. NGC 3201 and NGC 6101), \textit{and} clusters with full spatially mixed populations ($A^+ \sim 0$, e.g. NGC 288, NGC 4590, NGC 5053, NGC 5904, NGC 7078\footnote{See Section \ref{sec:Apluszero} for a detailed discussion on NGC 7078} and NGC 7089) in the same small dynamical age range (age/relaxation time $<$ 4.5). The central concentration of a primordial population seems to be in tension with the prediction of globular cluster formation models where P2 stars are preferentially concentrated towards the centre.

We also investigated the relationship between $A^+_{4}$ and the mass loss fraction ($\rm{M_c / M_{i}}$). Clusters that have lost >70\% of their initial masses due to dynamical evolution should be entirely mixed according to \cite{2013Vesperini}. However, given that their simulations do not include the effects of stellar evolution, our present day masses cannot be directly compared with \cite{2013Vesperini}. To do this we need to take into account that star clusters lose $\sim50\%$ of their mass during a Hubble time due to stellar evolution (e.g. high mass stars dying first), so the \cite{2013Vesperini} clusters that have lost >70\% of their initial mass correspond to the clusters with $\rm{M_c/M_i} \gtrsim 0.15$ in Figure \ref{fig:Aplusmassloss}. Therefore, clusters with $\rm{M_c/M_i} \gtrsim 0.15$ are giving us a peek into the diversity of configurations the P1 and P2 populations of stars in globular clusters can display at the time of birth. As expected, in Figure \ref{fig:Aplusmassloss} we found that the clusters with significant central concentrations in either P1 or P2 have undergone the least amount of mass loss, with the exception of NGC 6809. Generally, as more mass is lost by a cluster, the initial concentrations of the multiple populations are also lost, as the stars become spatially mixed. We therefore concentrate our analysis on the clusters that should have retained the largest amount of their initial conditions in terms of dynamical age and mass loss.

While 20 Galactic GCs were investigated by \cite{2019dalessandro}, our study overlaps with only 8 of these clusters. We tested for consistency with their results by matching the constraints of their analysis and found all 8 overlapping clusters produce the same cumulative radial distributions as \cite{2019dalessandro}. These constraints included limiting the HST field to 2 $R_{\rm{hlp}}$ in order to match the radial range covered by their analysis and only including the ground-based data for the analysis of NGC 288 within this same radial range. For our independent analysis, we included the ground-based photometry without restricting the radial range to 2 $R_{\rm{hlp}}$ and still found agreement with \cite{2019dalessandro} for 7 out of the 8 overlapping clusters, since both the HST and ground-based photometry show $A^+ \sim 0$. The one cluster that did not agree with their results is NGC 6101, where we found P1 to be centrally concentrated. When we considered only the HST photometry for NGC 6101, we found $A^+ \sim 0$ in agreement with \cite{2019dalessandro}, but with the inclusion of the ground-based photometry and therefore the outer region of the cluster, we found a P1 central concentration. This suggests that conclusions arrived at by studying only the inner regions of a cluster may be misleading, especially in dynamically young clusters. A more extensive coverage of such clusters is required to obtain a full picture.
 
Our results for the dynamically young clusters suggest that clusters are able to form with either enriched stars in the center, primordial stars in the center, or enriched and primordial stars distributed in the same way. This is an intriguing result, considering that the majority of globular cluster formation models will naturally produce clusters in which the P2 stars are centrally concentrated. Our results therefore argue for the need of additional theories that can explain how clusters form with mixed stellar populations or centrally concentrated primordial stars.

\subsection{Global Trends using Enriched Star Fractions ($\rm{P_2}$ / $\rm{P_{total}}$)}
\label{sec:pfracresults}

\begin{figure}
    \centering
    \includegraphics[width=\columnwidth]{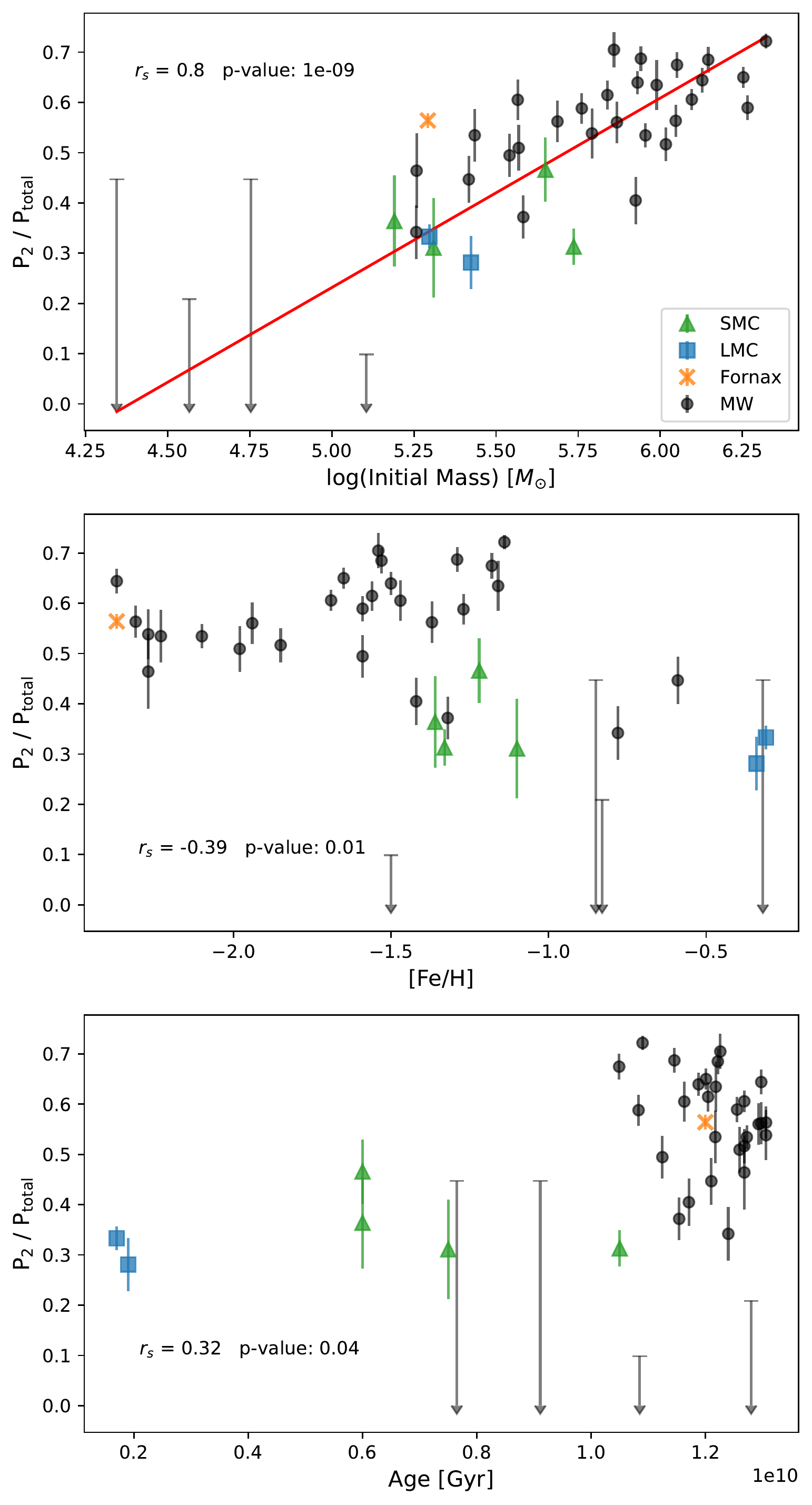}
    \caption{The enriched stellar population fraction as a function of global parameters for Galactic GCs (black circles). Added are SMC GCs (green triangles), LMC GCs (blue squares) and Fornax GCs (orange crosses). The large error bars plotted in grey for clusters with $\rm{P_2} / \rm{P_{total}} = 0$ are due to the low number of stars with spectroscopic abundance measurements. \textit{Top panel}: The fraction of P2 stars as a function of the initial mass of each cluster shows a clear correlation between the two parameters. \textit{Middle panel}: There is no significant relationship between the enriched star fraction and metallicity of the cluster. \textit{Bottom panel}:  There is no significant relationship between the enriched star fraction and the age of the cluster.}
    \label{fig:p2ratio}
\end{figure}

For each of the 28 Galactic GCs in our sample we calculated the enriched star fraction $\rm{P_2} / \rm{P_{total}}$ with associated standard errors, where enriched stars included both the P2 and P3 stellar populations. Unlike the cumulative radial distribution analysis, we did not implement a radial limit of 4.27$R_{\rm{hlp}}$ for each cluster, but instead calculated the $\rm{P_2} / \rm{P_{total}}$ fraction for the full possible extent of each cluster, taking into account the total completeness fraction (see Section \ref{sec:obsdata}). The top panel of Figure \ref{fig:p2ratio} shows the $\rm{P_2} / \rm{P_{total}}$ fraction as a function of the initial cluster mass. We obtain a strong correlation between these two parameters with $r_s = 0.8$ and p-value $ = 1\times10^{-9}$, similar to the correlation found by \cite{2017milone} and \cite{2019milone} using the $\rm{P_1} / \rm{P_{total}}$ fraction against log$(M[M_\odot])$. We found no significant correlations for the global fraction $\rm{P_2} / \rm{P_{total}}$ as a function of either metallicity or age (see Figure \ref{fig:p2ratio}). After removing the mass trend from our data, we similarly found that the residuals are uncorrelated with age or metallicity. We neither found significant correlations with orbital parameters such as peri- and apogalactic distances and eccentricity, nor with the slope of the mass function.

In order to test how young and low mass clusters fit into the global trends, we included an additional 7 Local Group clusters: NGC 121, NGC 336, NGC 416, NGC 1783, NGC 1978, Lindsay 1 and Fornax 3. The Local Group clusters were separated into multiple stellar populations using only HST photometry, but with the same method as outlined in Section \ref{sec:popsplit}. There is no need to combine HST and ground-based photometry for these clusters due to the fact that the half-light radius for each cluster is well within a single HST field, meaning the majority - if not all - stars are covered ny a single field. We only calculated the enriched star fractions $\rm{P_2} / \rm{P_{total}}$ for these additional clusters. In order to separate the populations, we used the narrow-band filter $F343N$, which contains the NH absorption line and can be used in the colour combination $C_{\rm{UBUn}} = (U - B) - (B - Un) = (F336W - F438W) - (F438W - F343N)$ as introduced by \cite{2017cunbi}. In the same way we confirmed consistency between the $\Delta C_{\rm{UBI}}$ distribution and chromosome maps, we also produced consistent results between the  $C_{\rm{UBUn}}$ and $\Delta C_{\rm{UBI}}$ distributions.

We also included an additional 4 low-mass Milky Way clusters to our sample (Ruprecht 106, Palomar 12, Terzan 7 and E3), using previous work which performed spectroscopic analysis of stars and found no evidence of multiple populations. E3 and Ruprecht 106 do not contain enriched populations according to the analysis of \cite{2018E3,2015E3} and \cite{2018Rup106P2,2021Rup106}, respectively, and we therefore set them to $\rm{P_2} / \rm{P_{total}} = 0$ with standard errors of $1 / \sqrt{N}$, where $N$ is the number of stars analysed. Similarly, the current consensus is that Terzan 7 and Palomar 12 do not contain multiple populations, based on the spectroscopic analysis of $\leq 5$ RGB stars \citep{2005Ter7,2004Pal12}, and we therefore set $\rm{P_2} / \rm{P_{total}} = 0$. The standard errors for the enriched star fraction associated with Terzan 7 and Palomar 12 were comparatively much larger than for other clusters, in order to reflect the uncertainty of declaring a non-detection of MPs with a sample of only 5 RGB stars. 
The age and metallicity of Lindsay 1 were taken from \cite{2009Lindsay1}, while those of E3 were taken from \cite{2010forbes}, and of Ruprecht 106 from \cite{2019Kruijssen}, who averaged the values determined by \cite{2010forbes} and \cite{2010dotter,2011dotter}. All other additional cluster ages and metallicities were taken from \cite{2019usher}.

The addition of these 11 young and low-mass GCs to the sample did not significantly influence the trends found for the $\rm{P_2} / \rm{P_{total}}$ fractions against global parameters. The initial mass correlation in Figure \ref{fig:p2ratio} is supported by the addition of these clusters, which continue the trend into the lower initial mass range. The relationship between $\rm{P_2} / \rm{P_{total}}$ and metallicity [Fe/H] previously showed a Spearman rank order coefficient of $r_s = 0.11$ and p-value $= 0.58$ for the original 28 Galactic GCs. After the addition of the 11 young and low-mass GCs, this coefficient changed to $r_s = -0.39$ with a p-value $= 0.01$, showing a slight but ultimately inconclusive anti-correlation. For $\rm{P_2} / \rm{P_{total}}$ against age, the Spearman correlation only changes from $r_s = -0.21$ with a p-value $= 0.29$ for the original 28 Galactic GCs, to $r_s = 0.32$ with a p-value $= 0.04$ for the full sample, again showing an inconclusive (weak) correlation. There appears to be no significant trend between enriched star fractions and metallicity or age, but the addition of a larger sample of young and low-mass clusters may alter this result.

\subsection{Dynamically Young Clusters}
\label{sec:notableclusters}

By `dynamically young' we refer to the clusters in our sample with dynamical ages $< 4.5$. \cite{2013Vesperini} found that dynamical age is a good indicator for the degree of dynamical mixing, with small dynamical ages corresponding to clusters which have retained the initial conditions of their formation. Following this criterion, the clusters described in detail throughout this section are assumed to have preserved their initial conditions. We have divided this section into three parts, focusing on dynamically young clusters with: enriched (P2) populations concentrated in the center in Section \ref{sec:P2conc}, the primordial (P1) population in the center in Section \ref{sec:P1conc} and spatially mixed populations in Section \ref{sec:Apluszero}. The cumulative radial distribution plots for the covered extent of all dynamically young clusters can be found in Appendix \ref{sec:Appendix A}.

\begin{table*}
\centering
\caption{Parameters for the 28 Galactic GCs studied in this paper. The individual columns give the final number of stars in our sample after the analysis of Sections \ref{sec:obsdata} and \ref{sec:popsplit}, split between the HST ($\rm{N_{HST}}$) and the ground-based ($\rm{N_{GB}}$) catalogues. The total cumulative radial distribution parameter $A^+_{4}$ was calculated for all clusters at a maximum radius of 4.27$R_{\rm{hlp}}$ in units of projected half-light radius, except for clusters specified in Section \ref{sec:Aplusresults}. We also include the $A^+_{\rm{total}}$ values calculated for the largest extent of each cluster covered by our datasets. The enriched star fractions $\rm{P_2 / P_{tot,4}}$ were also calculated for a radius of 4.27$R_{\rm{hlp}}$, and the full range ($\rm{P_2 / P_{tot}}$). The next column gives the maximum analysed radius in each cluster in units of projected half-light radius ($r_{\rm{max}} $ [HLR]). The final columns give the dynamical ages ($\rm{Age/T_{rh}}$), mass loss fractions ($\rm{M_c/M_i}$) and projected half-light radii ($R_{\rm{hlp}}$) (see Section \ref{sec:results} for details).}
\label{tab:results}
\resizebox{\textwidth}{!}{%
\begin{tabular}{cccccccccccc}
\hline
Cluster & $\rm{N_{HST}}$ & $\rm{N_{GB}}$ & $A^+_{4}$ & $A^+_{\rm{total}}$ & $\rm{P_2 / P_{tot,4}}$ & $\rm{P_2 / P_{tot}}$ & $r_{\rm{max}} $ [HLR] & Age / $\rm{T_{rh}}$ & $\rm{M_c/M_i}$ & $R_{\rm{hlp}}$ [pc] & Fe/H \\
\hline
NGC 288 & 190 & 356 & 0.21 $\pm$ 0.08 & 0.25 $\pm$ 0.09 & 0.37 $\pm$ 0.04 & 0.37 $\pm$ 0.04 & 4.61 & 3.18 $\pm$ 0.23 & 0.244 $\pm$ 0.007 & 5.83 & -1.32 \\
NGC 1261 & 903 & 153 & 0.02 $\pm$ 0.06 & 0.40 $\pm$ 0.15 & 0.58 $\pm$ 0.03 & 0.58 $\pm$ 0.03 & 25.74 & 5.68 $\pm$ 0.32 & 0.316 $\pm$ 0.005 & 3.25 & -1.27 \\
NGC 1851 & 1241 & 256 & -0.08 $\pm$ 0.07 & -0.94 $\pm$ 0.25 & 0.70 $\pm$ 0.03 & 0.67 $\pm$ 0.03 & 24.06 & 7.60 $\pm$ 0.43 & 0.283 $\pm$ 0.004 & 1.74 & -1.18\\
NGC 2808 & 3356 & 1401 & -0.08 $\pm$ 0.03 & -0.49 $\pm$ 0.08 & 0.75 $\pm$ 0.02 & 0.74 $\pm$ 0.01 & 20.86 & 3.61 $\pm$ 0.10 & 0.412 $\pm$ 0.003 & 2.45 & -1.14\\
NGC 3201 & 187 & 363 & 0.46 $\pm$ 0.12 & 0.46 $\pm$ 0.12 & 0.49 $\pm$ 0.04 & 0.51 $\pm$ 0.04 & 4.00 & 3.48 $\pm$ 0.24 & 0.462 $\pm$ 0.009 & 3.80 & -1.59\\
NGC 4590 & 217 & 144 & -0.13 $\pm$ 0.10 & 0.04 $\pm$ 0.15 & 0.53 $\pm$ 0.05 & 0.52 $\pm$ 0.05 & 5.86 & 2.73 $\pm$ 0.18 & 0.448 $\pm$ 0.036 & 4.44 & -2.23\\
NGC 4833 & 535 & 336 & -0.07 $\pm$ 0.07 & 0.02 $\pm$ 0.08 & 0.51 $\pm$ 0.03 & 0.52 $\pm$ 0.03 & 4.57 & 6.36 $\pm$ 0.28 & 0.199 $\pm$ 0.010 & 3.26 & -1.85 \\
NGC 5024 & 1293 & 438 & -0.42 $\pm$ 0.05 & -0.84 $\pm$ 0.11 & 0.56 $\pm$ 0.03 & 0.52 $\pm$ 0.02 & 10.31 & 1.33 $\pm$ 0.07 & 0.505 $\pm$ 0.036 & 6.43 &-2.10\\
NGC 5053 & 0 & 181 & 0.07 $\pm$ 0.16 & 0.07 $\pm$ 0.16 & 0.46 $\pm$ 0.08 & 0.45 $\pm$ 0.08 & 3.50 & 1.33 $\pm$ 0.16 & 0.410 $\pm$ 0.089 & 12.37 &-2.27\\
NGC 5272 & 1259 & 619 & -0.36 $\pm$ 0.05 & -0.17 $\pm$ 0.10 & 0.64 $\pm$ 0.02 & 0.64 $\pm$ 0.02 & 14.36 & 2.92 $\pm$ 0.14 & 0.476 $\pm$ 0.020 & 3.39 & -1.50\\
NGC 5286 & 1990 & 242 & 0.09 $\pm$ 0.05 & -0.23 $\pm$ 0.12 & 0.61 $\pm$ 0.02 & 0.59 $\pm$ 0.02 & 9.56 & 6.21 $\pm$ 0.25 & 0.283 $\pm$ 0.010 & 2.37 &-1.69\\
NGC 5904 & 970 & 657 & -0.10 $\pm$ 0.06 & -0.44 $\pm$ 0.15 & 0.70 $\pm$ 0.03 & 0.68 $\pm$ 0.02 & 14.28 & 3.23 $\pm$ 0.20 & 0.451 $\pm$ 0.008 & 3.51 &-1.29\\
NGC 5986 & 1278 & 329 & -0.08 $\pm$ 0.04 & -0.12 $\pm$ 0.05 & 0.59 $\pm$ 0.03 & 0.59 $\pm$ 0.02 & 5.99 & 7.06 $\pm$ 0.43 & 0.181 $\pm$ 0.013 & 2.77 &-1.59\\
NGC 6101 & 252 & 229 & 0.66 $\pm$ 0.13 & 0.70 $\pm$ 0.13 & 0.52 $\pm$ 0.05 & 0.53 $\pm$ 0.05 & 4.27 & 1.15 $\pm$ 0.11 & 0.479 $\pm$ 0.087 & 9.56 &-1.98\\
NGC 6121 & 197 & 208 & -0.08 $\pm$ 0.10 & -0.08 $\pm$ 0.10 & 0.64 $\pm$ 0.05 & 0.62 $\pm$ 0.05 & 2.89 & 15.69 $\pm$ 0.76 & 0.089 $\pm$ 0.001 & 2.49 &-1.16\\
NGC 6205 & 1093 & 421 & -0.04 $\pm$ 0.04 & -0.02 $\pm$ 0.07 & 0.68 $\pm$ 0.03 & 0.68 $\pm$ 0.03 & 10.19 & 3.86 $\pm$ 0.22 & 0.390 $\pm$ 0.015 & 3.46 &-1.53\\
NGC 6218 & 247 & 333 & 0.20 $\pm$ 0.07 & 0.12 $\pm$ 0.09 & 0.57 $\pm$ 0.04 & 0.55 $\pm$ 0.04 & 5.38 & 12.68 $\pm$ 0.43 & 0.220 $\pm$ 0.007 & 2.83 &-1.37\\
NGC 6254 & 649 & 524 & 0.02 $\pm$ 0.05 & -0.04 $\pm$ 0.09 & 0.62 $\pm$ 0.03 & 0.61 $\pm$ 0.03 & 6.82 & 5.90 $\pm$ 0.39 & 0.297 $\pm$ 0.006 & 2.96 &-1.56 \\
NGC 6341 & 740 & 238 & -0.16 $\pm$ 0.07 & -0.52 $\pm$ 0.23 & 0.57 $\pm$ 0.03 & 0.55 $\pm$ 0.03 & 20.29 & 5.32 $\pm$ 0.15 & 0.316 $\pm$ 0.004 & 2.39&-2.31 \\
NGC 6366 & 88 & 371 & 0.19 $\pm$ 0.14 & 0.09 $\pm$ 0.16 & 0.45 $\pm$ 0.05 & 0.47 $\pm$ 0.05 & 4.38 & 12.39 $\pm$ 1.11 & 0.144 $\pm$ 0.007 & 3.77 &-0.59\\
NGC 6752 & 437 & 376 & 0.08 $\pm$ 0.08 & -0.05 $\pm$ 0.14 & 0.71 $\pm$ 0.04 & 0.72 $\pm$ 0.04 & 9.59 & 4.88 $\pm$ 0.18 & 0.382 $\pm$ 0.005 & 2.87 &-1.54\\
NGC 6809 & 216 & 364 & -0.51 $\pm$ 0.07 & -0.49 $\pm$ 0.07 & 0.56 $\pm$ 0.04 & 0.56 $\pm$ 0.04 & 4.51 & 3.64 $\pm$ 0.18 & 0.261 $\pm$ 0.011 & 4.58 &-1.94 \\
NGC 6838 & 135 & 213 & 0.05 $\pm$ 0.13 & 0.01 $\pm$ 0.12 & 0.34 $\pm$ 0.06 & 0.34 $\pm$ 0.06 & 2.53 & 12.98 $\pm$ 1.27 & 0.253 $\pm$ 0.013 & 3.35 &-0.78\\
NGC 6934 & 499 & 119 & 0.21 $\pm$ 0.07 & 0.05 $\pm$ 0.14 & 0.61 $\pm$ 0.04 & 0.61 $\pm$ 0.04 & 8.87 & 6.39 $\pm$ 0.52 & 0.371 $\pm$ 0.052 & 2.95 &-1.47\\
NGC 6981 & 329 & 123 & 0.00 $\pm$ 0.09 & 0.19 $\pm$ 0.15 & 0.40 $\pm$ 0.05 & 0.40 $\pm$ 0.05 & 7.69 & 8.48 $\pm$ 0.98 & 0.082 $\pm$ 0.015 & 4.14 &-1.42 \\
NGC 7078 & 1352 & 272 & -0.03 $\pm$ 0.06 & 0.37 $\pm$ 0.10 & 0.62 $\pm$ 0.03 & 0.64 $\pm$ 0.03 & 15.00 & 4.10 $\pm$ 0.12 & 0.471 $\pm$ 0.005 & 2.03 &-2.37\\
NGC 7089 & 1815 & 422 & -0.06 $\pm$ 0.05 & -0.04 $\pm$ 0.16 & 0.64 $\pm$ 0.02 & 0.64 $\pm$ 0.02 & 24.35 & 3.54 $\pm$ 0.14 & 0.346 $\pm$ 0.006 & 3.04 &-1.65\\
NGC 7099 & 295 & 110 & 0.14 $\pm$ 0.09 & -0.11 $\pm$ 0.16 & 0.55 $\pm$ 0.05 & 0.54 $\pm$ 0.05 & 8.63 & 5.83 $\pm$ 0.20 & 0.231 $\pm$ 0.010 & 2.54 &-2.27\\ \hline
\end{tabular}%
}
\end{table*}

\subsubsection{Clusters with centrally concentrated P2 stars}
\label{sec:P2conc}
In this section we discuss the individual results of the clusters NGC 2808, NGC 5024, NGC 5272 and NGC 6809, which  contain a significant central concentration of the enriched (P2) stars.\\

NGC 2808 was separated into multiple stellar populations by \citet{Milone15} using a chromosome map with HST photometry. We find that the inner region covered by the HST field indicated that primordial and enriched stars are spatially mixed with $A^+ = 0.02 \pm 0.02$, whereas \cite{2019dalessandro} found $A^+ = -0.029 \pm 0.001$, in agreement with our results over the same approximate spatial range, i.e. 2 $R_{\rm{hlp}}$. However, the inclusion of stars in the ground-based photometry shows a significant P2 central concentration for the full range of the cluster, with $A^+_{\rm{total}} = -0.49 \pm 0.08$. Limiting the spatial range to 4.27$R_{\rm{hlp}}$ resulted in a value of $A^+_{4} = -0.08 \pm 0.03$, further strengthening the idea that omitting the outer stars from radially dependent analyses can hide the true properties of clusters. NGC 2808 contains the largest sample of stars from all 28 analysed clusters, with 4757 stars in total. It presents a good opportunity for obtaining substantial amounts of individual spectra for further analysis. The final sample of the cluster contained 1323 P1 stars and 3433 P2 stars, which exacerbates the mass budget problem, especially considering that NGC 2808 has a young dynamical age and should still retain $\rm{M_c/M_i} \sim 0.41$ of its initial mass.\\

A spectroscopic analysis of NGC 5024 was performed by \cite{2016boberg} for 53 RGB stars within 500 arcseconds of the cluster center, discovering a centrally concentrated enriched population. This agrees with our cumulative radial distribution of $A^+_{4} = -0.42 \pm 0.05$, which includes stars from the cluster center to 739 arcseconds. However, for the two different methods used by \cite{2016boberg}, they find $\rm{P_2} / \rm{P_{total}} \sim 0.3$, while our results for the total enriched fraction shows $\rm{P_2} / \rm{P_{total}} = 0.52 \pm 0.02$. Since \cite{2016boberg} only used RGB stars with magnitudes $V < 15.5$, while our analysis includes the full RGB of stars with magnitudes $V < 19.3$, we argue that our enriched star fraction includes a larger and more complete sample and is therefore more indicative of the enriched star fraction. We found NGC 5024 has the highest amount of remaining initial mass with $\rm{M_c/M_i} \sim 0.51$, along with one of the lowest dynamical ages, meaning its initial conditions should not have changed significantly over time. From Figure \ref{fig:Aplus} we see that the photometry from the inner region alone provides a different picture than the combination of HST and ground-based photometry,  supporting the idea that dynamical mixing of the populations affects the center of the cluster before the outer regions. Although the HST region contained 1293 stars and the ground-based photometry contained 438 stars, these outermost stars prove to be crucial in arriving at the full picture.\\

NGC 5272 was previously analysed by \cite{2019dalessandro}, who used a combination of HST photometry and Str\"{o}mgren photometry from \cite{2016massari}. Additionally, \cite{2011lardo} used SDSS photometry for RGB stars beyond 100 arcseconds from the cluster center. Both discovered a centrally concentrated enriched population, consistent with our cumulative radial distribution of $A^+_{4} = -0.36 \pm 0.05$ for stars within 4.27 $R_{\rm{hlp}}$. However, we found that extending to the full possible extent of the cluster returned a value of $A^+_{\rm{total}} = -0.17 \pm 0.10$, showing a less significant P2 central concentration overall. We found NGC 5272 has retained a high fraction of its initial mass, estimated to be close to $\rm{M_c/M_i} \sim 0.48$, so we consider NGC 5272 to also largely preserve its initial conditions. Our cumulative radial distribution for the HST photometry alone shows no dynamical mixing between the populations with $A^+ \sim 0$, but the ground-based photometry indicates the outer regions are not yet mixed.

\cite{2019rain} identified two populations in NGC 6809 based on 11 RGB stars using high resolution FLAMES/UVES spectra. Their spectroscopic identification of two populations is consistent with our photometric identification of two populations in both photometric data sets. We found a centrally concentrated enriched population in both the HST and ground-based photometry, which indicates a lack of dynamical mixing within the center of the cluster when compared with NGC 5024 and NGC 5272. Interestingly, NGC 6809 is dynamically young but has lost a significant amount of its initial mass, with $\rm{M_c/M_i} \sim 0.26$. NGC 6809 has the smallest galactocentric distance in our sample, with $R_{GC} = 4.01 \pm 0.03$ kpc and an escape velocity of $v_{esc} = 17.3$ km/s. Tidal disruption affects clusters with smaller galactocentric distances more strongly \citep{2019baumgardthilker} and the size of an accreted cluster in particular will respond to the tidal field of the MW upon accretion \citep{2014miholics}. As NGC 6809 is both suggested to be an accreted cluster \citep{2019massari} and has a small galactocentric distance and relatively low escape velocity, we expect that although the cluster is dynamically young, tidal disruption after its accretion has affected its initial conditions. It therefore becomes somewhat difficult to confidently conclude whether our discovery of a centrally concentrated P2 population is representative of its initial spatial distribution.

\subsubsection{Clusters with centrally concentrated P1 stars}
\label{sec:P1conc}

One of the most interesting results of this work is the centrally concentrated primordial populations found in NGC 3201 and NGC 6101. In order to test the validity of these findings, we present a more thorough analysis of the two clusters in this section.\\

NGC 3201 is considered dynamically young, but previous studies of the cluster have produced complicated results that cause uncertainty around whether we can assume it maintains its initial configuration. NGC 3201 is proposed to be an accreted cluster previously belonging to Sequoia/Gaia-Enceladus \citep{2019massari}. \cite{2015lucatello} found that the P1 population in NGC 3201 hosts a higher fraction of binary stars than the P2 population, which they suggested to be due to the dense conditions of the central region that enhance the destruction and ejection of binaries. This result assumes that only P2 stars can be centrally concentrated. \cite{2020kamann} used HST photometry and MUSE spectroscopy and also found that NGC 3201 contains a higher binary fraction in the P1 population than it does for P2. They compare this result to simulations suggesting P1 binaries are only overabundant outside the half-light radius \citep{2015hong,2016hong}. These simulations also assume a P2 central concentration, as they use this configuration for the initial conditions of their simulation. Our discovery of a P1 concentration ($A^+_{4} = 0.46 \pm 0.12$) therefore does not support the previous hypothesis proposed to describe the relative binary fractions between different sub-populations, but our result is not unique in that \cite{2022Hartmann} also discovered a P1 central concentration by combining HST photometry with photometry from the S-PLUS survey. \cite{2019bianchini} and \cite{2021wan} investigated the peculiar kinematics in the outskirts of NGC 3201, which contains tidal tails and exhibits flattened velocity dispersions in the outskirts.

When analysing NGC 3201, we found that it suffered from significant differential reddening. However, after correcting for its effect (see Section \ref{sec:reddening}), the final spatial distribution of the populations showed no indication of problems due to differential reddening. In NGC 3201, we found that P1 stars had the highest concentration at intermediate radii around 150’’, with P2 stars being dominant in the outer parts and also towards the center of the cluster. A KS test showed that the central concentration of P2
was significant at a $\sim 2 \sigma$ level and significant at the $8 \sigma$ level towards the
outer parts, leading to a U-shaped distribution in the relative fraction of P2 stars.

In order to properly test the validity of the primordial central concentration discovery in the $A^+$ parameter, we performed the probability cut and population limit tests outlined at the end of Section \ref{sec:GMM}. By testing the effect of different limits in $\Delta C_{\rm{UBI}}$ to separate the populations, we found $A^+ = 0.28 \pm 0.25$. Similarly, by testing different probability thresholds for the membership of stars belonging to P1 and P2, we found $A^+ = 0.35 \pm 0.03$. These tests confirm that the presence of a centrally concentrated primordial population is a consistent/robust result regardless of the method chosen to classify P1/P2 stars.
Our discovery of a centrally concentrated primordial population could indicate that the peculiar kinematics found by \cite{2019bianchini} and \cite{2021wan} is driven by the enriched population of stars in the outskirts. NGC 3201 has intriguing characteristics and our discovery of a P1 central concentration further adds to these previous results. However, it is difficult to describe the complexity of NGC 3201 using only the $A^+$ parameter and future work would benefit from a parameter which incorporates both the radial spatial distributions between populations and the enriched star fraction for such clusters.\\

\cite{2019dalessandro} analysed NGC 6101 and found $A^+ = -0.003 \pm 0.001$, indicating the populations are homogeneously mixed. In our analysis we found $A^+ = -0.07 \pm 0.02$ for 252 stars in the HST photometry, whereas $A^+ = 0.57 \pm 0.19$ was found using 229 stars in the ground-based photometry alone. Our combined cumulative radial distributions indicate a centrally concentrated primordial population. NGC 6101 is the only case in our sample for which the HST and ground-based separations using the $\Delta C_{\rm{UBI}}$ distributions returned a different number of populations. The chromosome map returned two populations, as did the $\Delta C_{\rm{UBI}}$ distribution for the HST photometry. However, in the ground-based  $\Delta C_{\rm{UBI}}$ distribution, three populations were returned. The blending of the populations was also somewhat present in the chromosome map, but two populations are nonetheless distinct enough for separation, as is also shown in Figure 7 of \cite{2017milone} where the primordial population contains more stars than the enriched population. We found NGC 6101 has retained almost half of its initial mass ($\rm{M_c/M_i} \sim 0.48$) and has gone through the least amount of dynamical mixing of all 28 clusters. With a low metallicity of [Fe/H]$=-$1.98 dex \citep{2010harris}, the populations in a $\Delta C_{\rm{UBI}}$ distribution are closer together than in more metal rich targets, since $C_{\rm{UBI}}$ is most sensitive to molecular bands, which are weaker at low metallicities. This leads to difficulties in separating the populations. Due to that, we thoroughly tested how the separation of populations affected the final cumulative radial distributions. The result of trying different probability thresholds for the memberships of stars belonging to P1 and P2 returned a value of $A^+ = 0.59 \pm 0.06$, while the test of sampling arbitrary limits in the $\Delta C_{\rm{UBI}}$ colour distributions returned $A^+ = 0.39 \pm 0.19$, showing a robust signal that P1 is concentrated in all cases.\\

Some simulations have studied the concept of an initially centrally concentrated population evolving over time. For example, the simulations of \cite{2013Vesperini} show that for a dynamically young cluster with an initial P2 central concentration, the P2 fraction as a function of radius will decrease significantly in the outer regions of the cluster, due to the slowing of two-body relaxation at larger distances from the cluster centre. However, we note that the same could be concluded if P1 were to have been formed more centrally concentrated, as there is no physical distinction between stars labeled P1 or P2 in these simulations, other than their initial configurations. Therefore, the behaviour we observe from the dynamically young clusters in our sample is indicative of the initial conditions, where the P1 population was born initially more centrally concentrated.

When viewing only the inner HST region ($r < 1 R_{\rm{hlp}}$) of NGC 3201 (Figure \ref{fig:Aplus3201}), we found that the $\rm{P_2 / P_{total}}$ fraction decreases with increasing radius. We performed a K-S test on the P1 and P2 distributions within this range to quantify this, based on the standard two-sample test described in Section 12.4 of \cite{2001Monahan}, but modified to also include the weights ($w = 1/f_T$) of each star, following the method described in Equations 3-5 of \cite{2022Baumgardt1}. The weighted K-S test showed the P1 and P2 distributions have a 2\% probability of following the same distribution, meaning there is likely a P2 central concentration for the inner region. However, if we consider stars beyond $1 R_{\rm{hlp}}$ the enriched star fraction increases for the outer regions. Figure 7 of \cite{2013Vesperini} shows a simulated scenario in which the enriched star fraction as a function of radius could demonstrate similar U-shaped behaviour, however, it is not immediately clear that this represents the same phenomenon observed in NGC 3201.

For example, the radius at which \cite{2013Vesperini} expects this increase ($r > 5 R_{\rm{hlp}}$) is much larger than the radius at which we observe the increase ($r \sim 1 R_{\rm{hlp}}$). Moreover, the dynamical ages ($\rm{Age/T_{rh}}$) at which the U-shaped behaviour occurs in the simulations is expected to be $\rm{Age/T_{rh}} \geq 5$, whereas NGC 3201 has a dynamical age of $\rm{Age/T_{rh}} = 3.48 \pm 0.24$. Finally, \cite{2013Vesperini} describes this increase as a "weak final rise" on the order of $\sim10\%$, whereas in NGC 3201 we observe an $\sim 300 \%$ increase at an $\sim 8 \sigma$ significance between the minimum at $\sim 1 R_{\rm{hlp}}$ and the maximum at $\sim 4 R_{\rm{hlp}}$ of the enriched star fraction. Detailed simulations will be necessary to test how the initial conditions of NGC 3201 looked.

\subsubsection{Spatially mixed populations}
\label{sec:Apluszero}
We focus in this section on the dynamically young clusters that have retained most of their initial conditions but are nevertheless spatially mixed and do not contain one centrally concentrated population. These clusters include NGC 288, NGC 4590, NGC 5053, NGC 5904, NGC 6205, NGC 7078 and NGC 7089.\\

NGC 288 was analysed by \cite{2019dalessandro} using HST photometry, in which they found that it contains spatially mixed populations with $A^+ = -0.045 \pm 0.002$. Similarly, we found two spatially mixed populations, with $A^+_{4} = 0.21 \pm 0.08$ (P1 centrally concentrated only at $<3\sigma$ level). Additionally, \cite{2022Hartmann} used both, HST photometry and photometry from the S-PLUS survey, calculating cumulative radial distributions that show mixed populations in the central HST regions, but with a P2 central concentration in the outer regions. The discrepancies between our results in the outer regions - aside from the use of different photometric bands - appears to be due to differences in our analysis methods. More specifically, our sample of stars are corrected for photometric incompleteness, we exclude stars from our analysis in which the P1/P2 classifications are ambiguous ($p>80\%$), our limiting radius is 4.27 $R_{hlp}$ compared to their 5.5 $R_{hlp}$ and our sample includes an extra 116 stars in comparison. We found that NGC 288 has retained only a fraction $\rm{M_c/M_i} \sim 0.24$ of its initial mass, with an enriched fraction of $\rm{P_2} / \rm{P_{total}} = 0.37 \pm 0.04$. At a glance, it seems plausible that mass loss is responsible for ejecting either primordial or enriched stars from the outer regions, resulting in spatially mixed populations. However, it is also possible that NGC 288 formed with spatially mixed populations, as the initial configuration is difficult to determine due to the significant amount of mass loss.\\

We found that NGC 4590 contains spatially mixed populations for both the HST and ground-based photometry, but based on a comparatively small sample size of 361 stars. \cite{2019baumgardthilker} found that NGC 4590 has large perigalactic ($8.95 \pm 0.06$ kpc) and apogalactic ($29.51 \pm 0.42$ kpc) distances, and \cite{2019massari} suggests one of the Helmi streams is the progenitor of this cluster. We found NGC 4590 retains approximately $\rm{M_c/M_i} \sim 0.45$ of its initial mass and is one of the dynamically youngest clusters in our sample, but nonetheless contains fully spatially mixed populations. The large peri- and apogalactic distances suggest tidal stripping is unlikely to have removed a significant fraction of stars, but the accretion of NGC 4590 to the MW may have led to a stronger than predicted mass loss.\\

Previous work has found NGC 5053 to be dynamically complicated: it contains significant tidal tails \citep{2010jordi,2006lauchner} and a possible tidal bridge to NGC 5024 \citep{2010chun}. Although NGC 5053 has one of the lowest dynamical ages and is predicted to retain a significant fraction of its initial mass with $\rm{M_c} / \rm{M_{i}} \sim 0.41 $, we found its stellar populations are spatially mixed. NGC 5053 was the only cluster for which we relied solely on the ground-based photometry. Due to the insufficient number of RGB stars in the HST photometry, the full extent of the ground-based photometry - including the cluster center - was used instead. The core of NGC 5053 has the lowest density of any cluster in our sample, and it has a large half-light radius, greatly reducing the blending effect in the cluster center that usually plagues ground-based photometry. As it is possible that NGC 5053 and NGC 5024 were accreted together within the same dwarf galaxy, we note that this event may have affected the mass loss of both clusters. \\

The work of \cite{2019Lee} using Str\"{o}mgren photometry and the $\rm{C_{UBI}}$ index found two populations in NGC 5904 with spatially mixed populations. In a follow-up paper, \cite{2021lee} stated that this previously determined bimodal distribution could actually be separated further into three populations using Str\"{o}mgren and Ca-CN-CH-NH photometry. With this difference in classifications, their cumulative radial distributions changed from showing spatially mixed populations throughout the extent of the cluster - consistent with our results - to instead showing the most carbon-poor and nitrogen-rich population as centrally concentrated. \cite{2011lardo} also separated NGC 5904 into two populations using SDSS photometry, which they refer to as UV-blue and UV-red. The resulting cumulative radial distributions from \cite{2011lardo} show the UV-red stars are more centrally concentrated. Our final sample of NGC 5904 contains a large sample size of 1627 RGB stars and was consistent between the HST and ground-based photometry in identifying two stellar populations, exhibiting complete spatial mixing between populations and a consistent enriched fraction of $\rm{P_2} / \rm{P_{total}} = 0.68 \pm 0.02$ throughout the cluster. We found that our results are consistent with only the initial findings of \cite{2019Lee}, as we did not find three populations within NGC 5904 using the combined HST and ground-based photometry. The introduction of spectroscopy to classify the populations based on chemical abundances such as carbon and nitrogen may help to check the validity of our photometrically separated populations.\\

NGC 6205 was found to have a mass loss ratio close to $\rm{M_c} / \rm{M_{i}} \sim 0.39$ and is spatially mixed to its outermost regions at $10.19 R_{hlp}$. Similarly, we found NGC 7089 has a mass loss ratio of $\rm{M_c} / \rm{M_{i}} \sim 0.35$ with spatially mixed populations extending out to $24 R_{hlp}$. Both clusters have large masses and are close to the upper limit of our definition of `dynamically young'. NGC 6205 has previously been analysed by \cite{2018savino} using both HST and Str\"{o}mgren photometry, in which they estimate an enriched fraction of approximately 80\%, compared to our fraction of $\rm{P_2} / \rm{P_{total}} = 0.68 \pm 0.03$. In terms of cumulative radial distributions, they also found no evidence for a centrally concentrated population in both the inner and outer regions of NGC 6205 (extending to approximately 700 arcseconds). NGC 7089 was analysed by \cite{2022Hartmann} using HST and S-PLUS survey photometry, discovering a P2 central concentration in both the HST field and outer region, which is at odds with our results of spatially mixed populations throughout the cluster. We also find our results at odds with \cite{2011lardo}, who identified a centrally concentrated population using cumulative radial distributions from SDSS photometry for both NGC 6205 and NGC 7089. If the dynamical age of NGC 6205 is long enough for dynamical mixing to occur throughout the entire cluster, we would expect this to occur for NGC 7078 and NGC 6809 as well, as per Figure \ref{fig:Aplustr}. However, we found clusters with similar dynamical ages have strongly varying spatial concentrations instead.\\

Previous photometric analysis of NGC 7078 has found contradictory results: \cite{2015larsen} combined HST and SDSS photometry of RGB stars and discovered three stellar populations, which yielded a centrally concentrated primordial population; however, \cite{2011lardo} found only two populations using SDSS photometry and consequentially discovered a centrally concentrated enriched population instead. The $m_{F336W}$ vs. $C_{F275W,F336W,F438W}$ plot of \cite{2015piotto} (Figure 22) shows at least two populations within NGC 7078 using HST photometry, while \cite{2017milone} distinctly separated the HST photometry into three populations using a chromosome map.\\
We found NGC 7078 contained one of the largest discrepancies for the P1 and P2 populations between the chromosome map and the $\Delta C_{\rm{UBI}}$ distribution, with a contamination of approximately 20\%. The low metallicity of NGC 7078 makes it difficult to separate the populations in the $\Delta C_{\rm{UBI}}$ distribution, as the molecular bands responsible for the colour variations in $\Delta C_{\rm{UBI}}$ become weaker, translating into smaller colour differences \citep[see discussion in e.g.][and references therein]{Balbinot22}. Additionally, this cluster suffers severely from differential reddening, which adds noise to the signal of the multiple populations. Taking these caveats into account, we advise the reader to take the following results for NGC 7078 with caution. We checked other low metallicity clusters in our sample ([Fe/H] $< -1.8$) and found they did not suffer from this same confusion, and NGC 7078 is the only cluster in our sample affected by this.\\
Because of the significant overlap between the Gaussians fitted by GMM to separate the populations in the $\Delta C_{\rm{UBI}}$ distribution for NGC 7078, we took a different approach. In the HST region, we rely on the chromosome map classification of populations, finding a resulting $A^+$ value close to zero, which indicates the centre of the cluster is spatially mixed. Guided by the HST data, we establish colour cuts for the ground-based data which gave us relatively pure P1 and P2 stars. More specifically, we selected only the extremes of the P1 ($\Delta C_{\rm{}UBI}<-0.7$) and P2 ($\Delta C_{\rm{}UBI}>-0.3$) populations. For reference we cross-matched our RGB stars with APOGEE DR17 \citep{2017apogee2,2022apogee1}, where the stars in our final sample correspond to [Al/Fe] abundances of [Al/Fe]$<0.05$ for P1 stars and [Al/Fe]$>0.4$ for P2 stars. The final $A^+_4$ and $A^+_{total}$ values quoted are therefore the combination of chromosome map classifications for the HST stars and our sample of extreme P1 and P2 stars selected as described above for the ground-based stars. We find $A^+_4 = -0.03 \pm 0.06$, indicative of spatially mixed populations out to $4.27 R_{\rm{hlp}}$, but with an overall signal indicating a P1 central concentration for the full extent (out to $15 R_{\rm{hlp}}$) of the cluster ($A^+_{\rm{total}}= 0.37 \pm 0.10$).\\

\subsection{Constraints on the loss of P1 stars}

Previous mass-loss scenarios involving internal enrichment aim to solve the mass budget problem by suggesting P1 stars are primarily located in the outskirts of GCs during formation, e.g. \cite{2020krause}. With P2 stars concentrated in the centre, mass-loss in the outskirts would then be responsible for the removal of P1 stars from the clusters \citep[e.g.][]{D08,Vesperini10}. \cite{2015bastian} explored this concept by analysing the correlations between enriched star fractions and cluster properties such as mass, metallicity and Galactocentric distance using literature data from 33 GCs.
For scenarios in which self-enrichment is responsible for the MP phenomenon, the enriched star fraction is expected to vary from the initial birth of the cluster to the present day, but was instead found to be constant throughout time, within errors. They concluded that the mass budget problem cannot be solved by assuming mass-loss in the outskirts of clusters, claiming that alternative theories are needed instead. \cite{2019gratton} suggested a combination of polluting and diluting scenarios may explain the resulting chemical abundance spreads observed in GCs, with an emphasis that the interacting binaries theory \citep{2012vanbeveren} may be responsible for the ejection of stars in clusters. According to their relative spatial distribution (i.e. $A^+$), we have found varying behaviours for the initial spatial configurations of MPs in our sample of Galactic GCs, where dynamically young clusters in our sample show P1 centrally concentrated stars, as well as a homogeneous mix of populations. However, this by itself does not necessarily translate to the exacerbation of the mass budget problem as one also needs to account for the relative number of P1 stars in the outskirts of the clusters (i.e. where stars more likely to escape from the cluster reside).

Our analysis has revealed that in the outer regions P1 stars do not constitute the majority of the stars, with the exception of NGC 5024 and NGC 6809 (see bottom right panels of Figure \ref{fig:Aplus} and Figures in Appendix \ref{sec:Appendix A}). This suggests that contrary to what is required by different models, during the dynamical evolution of these clusters P2 stars would be lost to the field population at a similar or higher rate than P1 stars. This would have important implications on the interpretations of the number of P2 stars found in the field, and their use to anchor the contribution of dissolved GCs to their host galaxy mass.

\section{Conclusions}

We have performed a spatially complete analysis of a large and diverse sample of 28 Galactic GCs, showing that GCs which still maintain their initial conditions can contain a central concentration of enriched \textit{or} primordial stars, as well as a homogeneous mix of both. We found centrally concentrated enriched populations in NGC 2808, NGC 5024, NGC 5272 and NGC 6809. They can be explained with existing formation theories that involve internal polluters, such as SMS, FRMS or AGB stars. We can also rely on the notion of dynamical mixing to explain why GCs with large dynamical ages tend to have spatially homogeneous stellar populations over time. However, dynamically young GCs with a centrally concentrated \textit{primordial} population (NGC 3201 and NGC 6101) cannot be explained with current formation theories. These models cannot account either for dynamically young GCs that already contain fully spatially mixed stellar populations such as NGC 288, NGC 4590, NGC 5053, NGC 5904, NGC 6205, NGC 7078 and NGC 7089. Furthermore, the existence of dynamically young clusters with fully mixed populations or a centrally concentrated P1, pose more challenges if P1 stars are required to be preferentially lost during the long term dynamical evolution of the cluster.

Interpolations or simulations based off an incomplete view of clusters have previously been used to constrain the possible fractions of primordial or enriched stars. In our analysis, we used a spatially complete view of each cluster to calculate the enriched star fractions ($\rm{P_2}$ / $\rm{P_{total}}$), which showed a clear correlation with the initial mass, but no clear correlations against other global parameters such as age and metallicity. Our sample of 28 Galactic GCs, 4 low-mass Galactic GCs and 7 Local Group GCs provided a range of $0 \leq \rm{P_2}$ / $\rm{P_{total}} < 0.75$ for the total enriched star fractions. We found that in some clusters, the enriched star fraction as a function of radius was constant across the extent of the cluster, while others exhibited either increasing or decreasing enriched star fractions.

Current theories of GC formation and theoretical simulations have assumed the possibility of only a P2 central concentration, due in part to an analysis which limits itself to only the central regions of clusters and assumes conclusions on the properties of the full cluster. We argue the need for future theories and simulations to also consider alternative configurations of initial conditions. The next stage of this research will explore the spectroscopic data available for our sample of 28 GCs in the same manner: combining data for the inner and outer regions of each cluster for a spatially complete view. We aim to check the validity of our photometric separations by spectroscopically separating the stellar populations based on chemical abundances. We will also use our current classifications of populations to explore the kinematic differences, along with differences in chemical abundances and binary fractions in order to provide further observational information relating to the possible initial conditions and the final, dynamically mixed conditions of the clusters in our sample. 

\section*{Acknowledgements}

We thank the referee for their insightful feedback on the manuscript, which improved the quality of our work.\\
The authors are very grateful to Florian Niederhofer for providing us an independent, HST based reddening map of NGC 7078.\\
This study was supported by the Klaus Tschira Foundation.

\section*{Data Availability}
The \textit{Hubble Space Telescope} UV Globular Cluster Survey (``HUGS'') photometric catalogue: \url{https://archive.stsci.edu/prepds/hugs/}\\
\\
The wide-field, ground-based Johnson-Cousins \textit{UBVRI} photometric catalogue, courtesy of Stetson et al. (2019), is available through the Canadian Astronomy Data center: \url{https://www.cadc-ccda.hia-iha.nrc-cnrc.gc.ca/en/community/STETSON/}\\
\\
The Gaia Early Data Release 3 (EDR3) archive: \url{https://gea.esac.esa.int/archive/}\\
\\
The Galactic Globular Cluster Database Version 2: \url{https://people.smp.uq.edu.au/HolgerBaumgardt/globular/}\\
\\
Metallicities from the Catalogue of Parameters for Milky Way Globular Clusters: The Database: \url{https://physics.mcmaster.ca/~harris/mwgc.dat}\\
\\
The Catalogue of galactic globular cluster surface brightness profiles courtesy of Trager et al. (1995) is available through the Strasbourg astronomical Data Center (CDS): \url{https://cdsarc.cds.unistra.fr/ftp/J/AJ/109/218/tables.dat}\\




\bibliographystyle{mnras}
\bibliography{example} 



\appendix
\section{Cumulative radial distributions of dynamically young clusters}
\label{sec:Appendix A}

\begin{figure*}
    \centering
    \includegraphics[width=\textwidth]{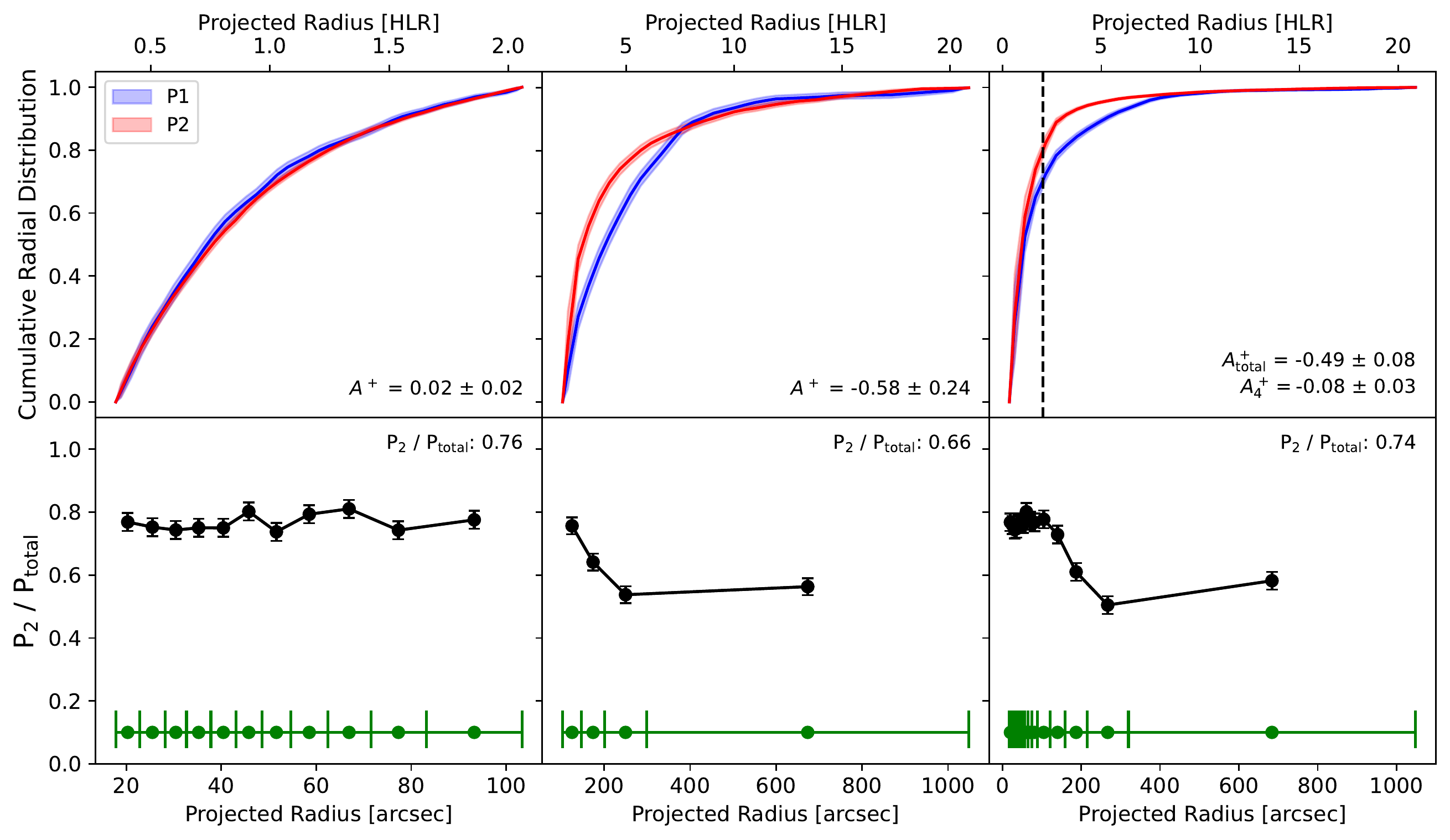}
    \caption{As in Figure \ref{fig:Aplus}, but for NGC 2808.}
    \label{fig:Aplus2808}
\end{figure*}

\begin{figure*}
    \centering
    \includegraphics[width=\textwidth]{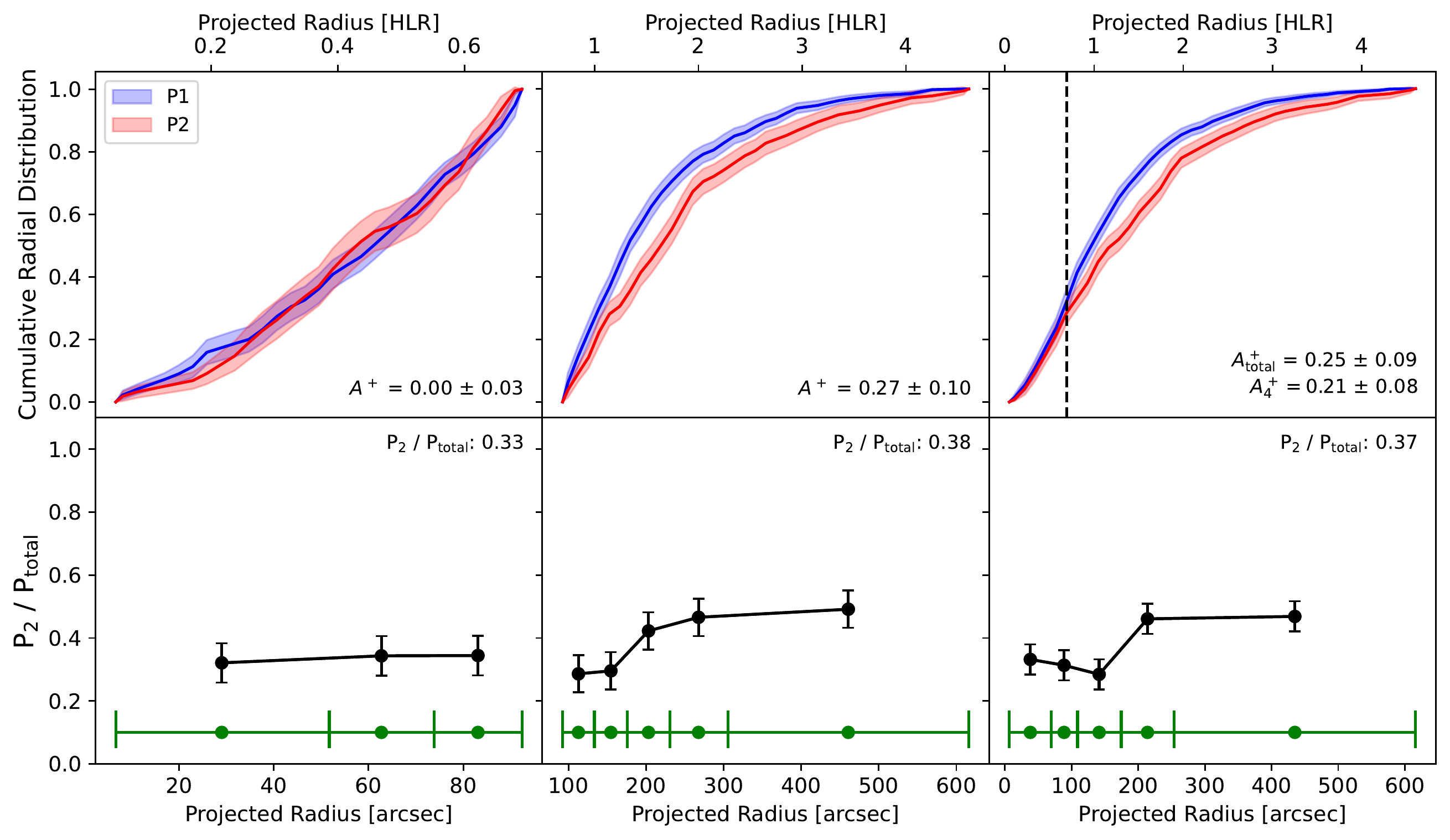}
    \caption{As in Figure \ref{fig:Aplus}, but for NGC 288.}
    \label{fig:Aplus288}
\end{figure*}

\begin{figure*}
    \centering
    \includegraphics[width=\textwidth]{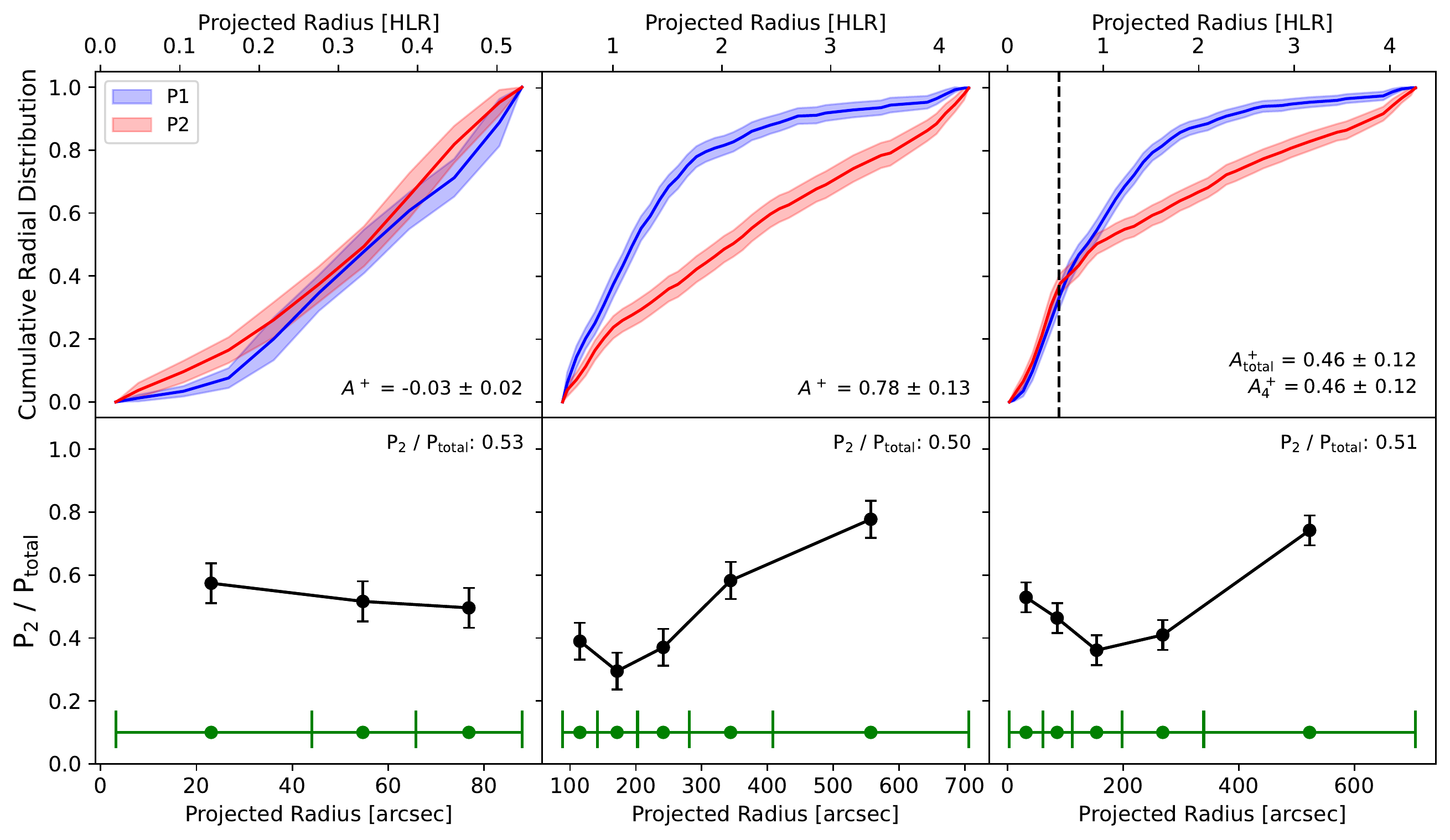}
    \caption{As in Figure \ref{fig:Aplus}, but for NGC 3201.}
    \label{fig:Aplus3201}
\end{figure*}

\begin{figure*}
    \centering
    \includegraphics[width=\textwidth]{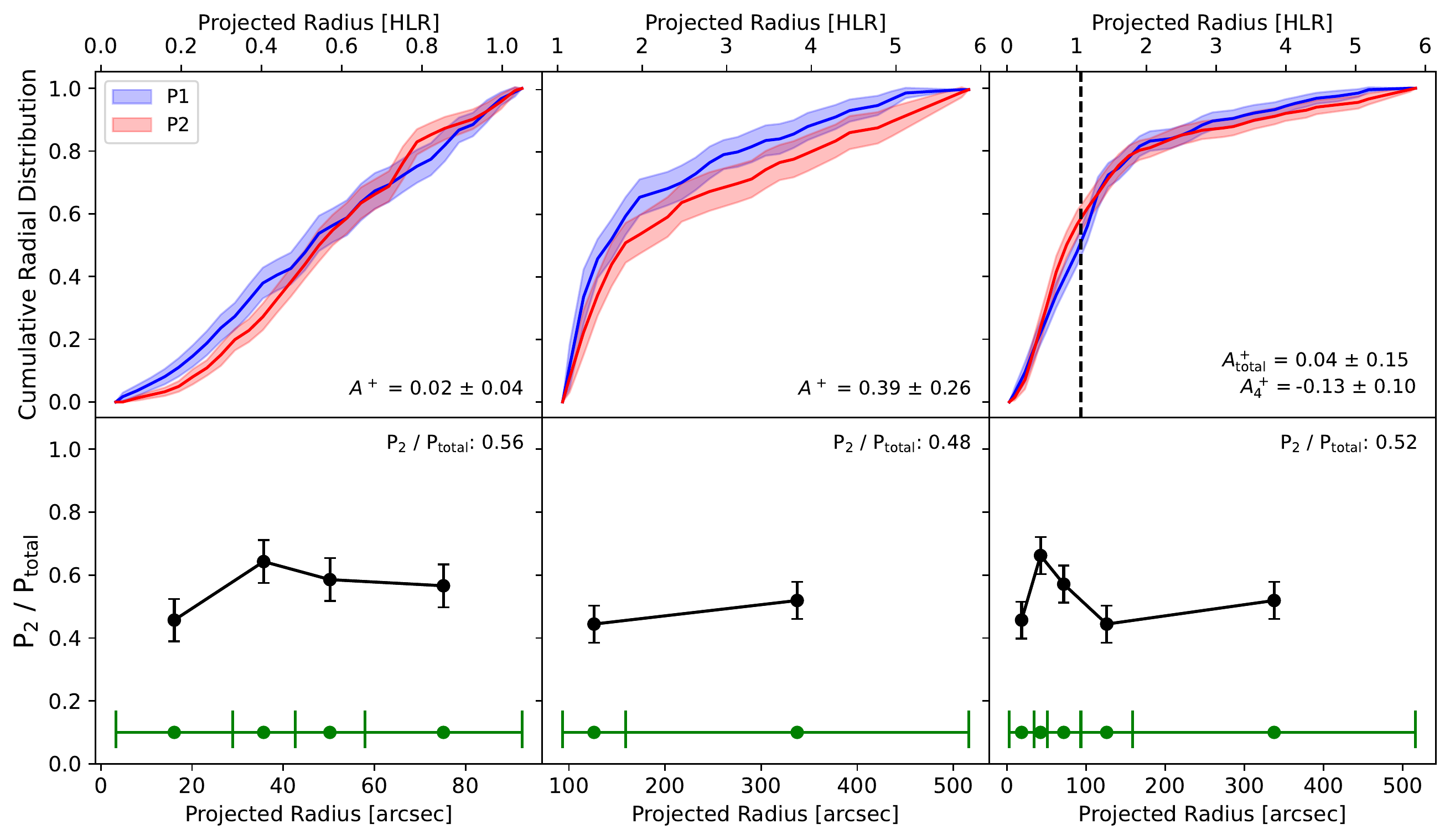}
    \caption{As in Figure \ref{fig:Aplus}, but for NGC 4590.}
    \label{fig:Aplus4590}
\end{figure*}

\begin{figure*}
    \centering
    \includegraphics[width=\textwidth]{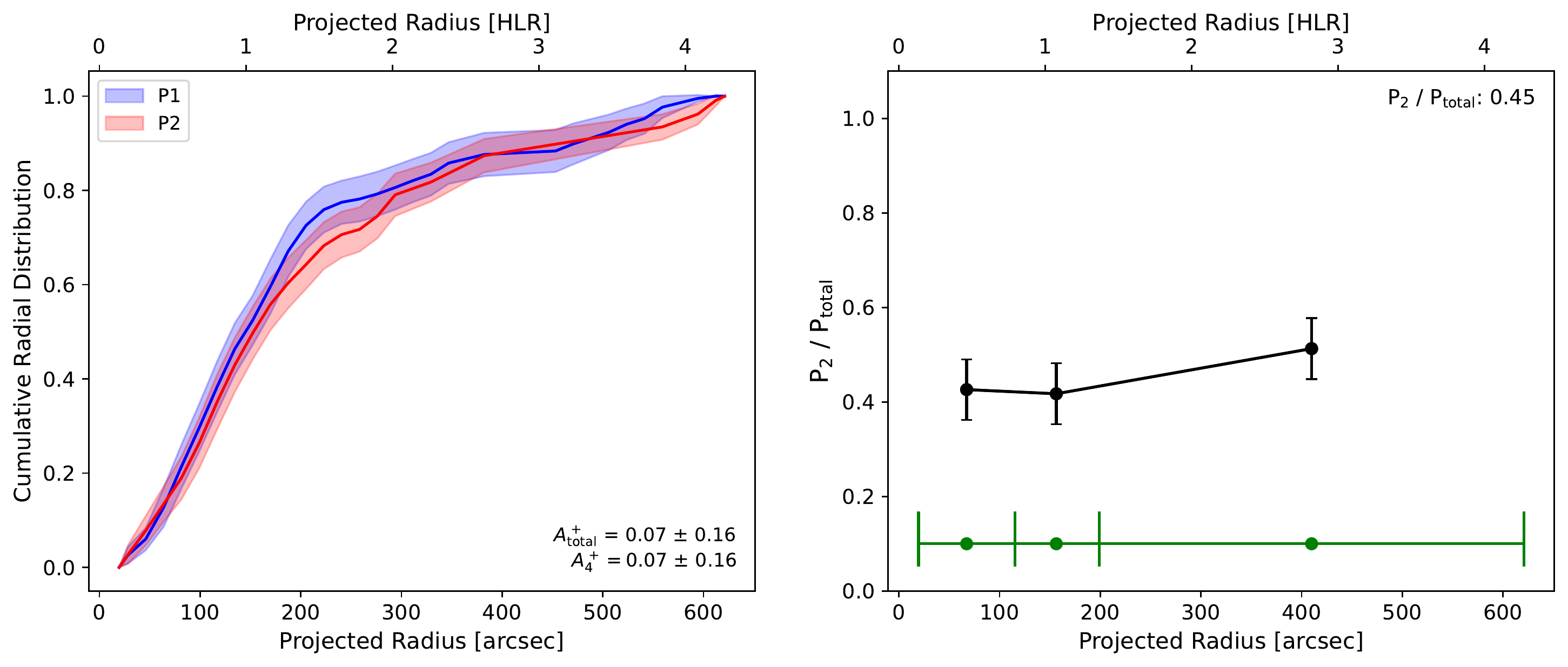}
    \caption{As in Figure \ref{fig:Aplus}, but for NGC 5053. No HST photometry was used, so only the ground-based photometry was included in the final sample.}
    \label{fig:Aplus5053}
\end{figure*}

\begin{figure*}
    \centering
    \includegraphics[width=\textwidth]{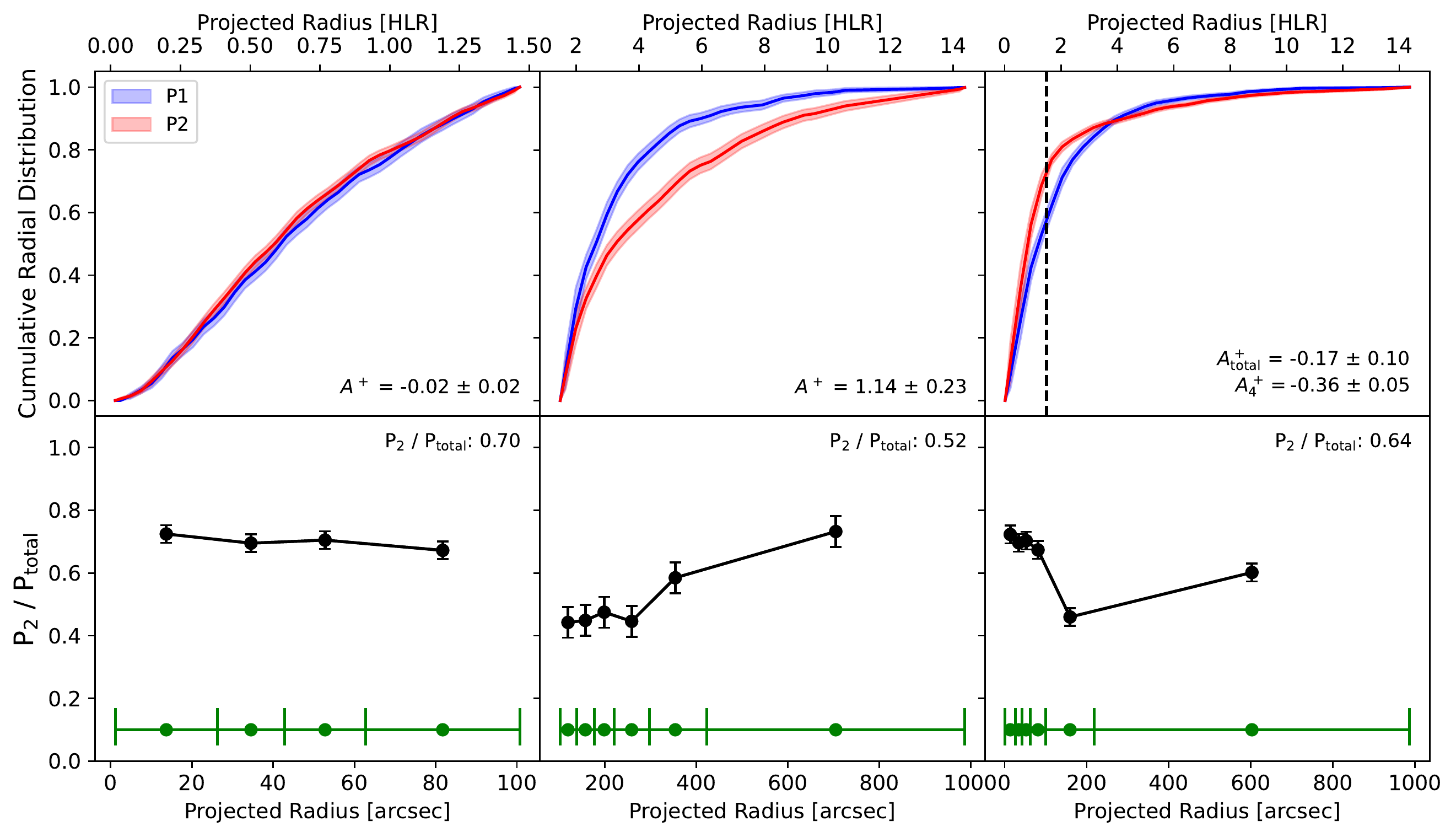}
    \caption{As in Figure \ref{fig:Aplus}, but for NGC 5272.}
    \label{fig:Aplus5272}
\end{figure*}

\begin{figure*}
    \centering
    \includegraphics[width=\textwidth]{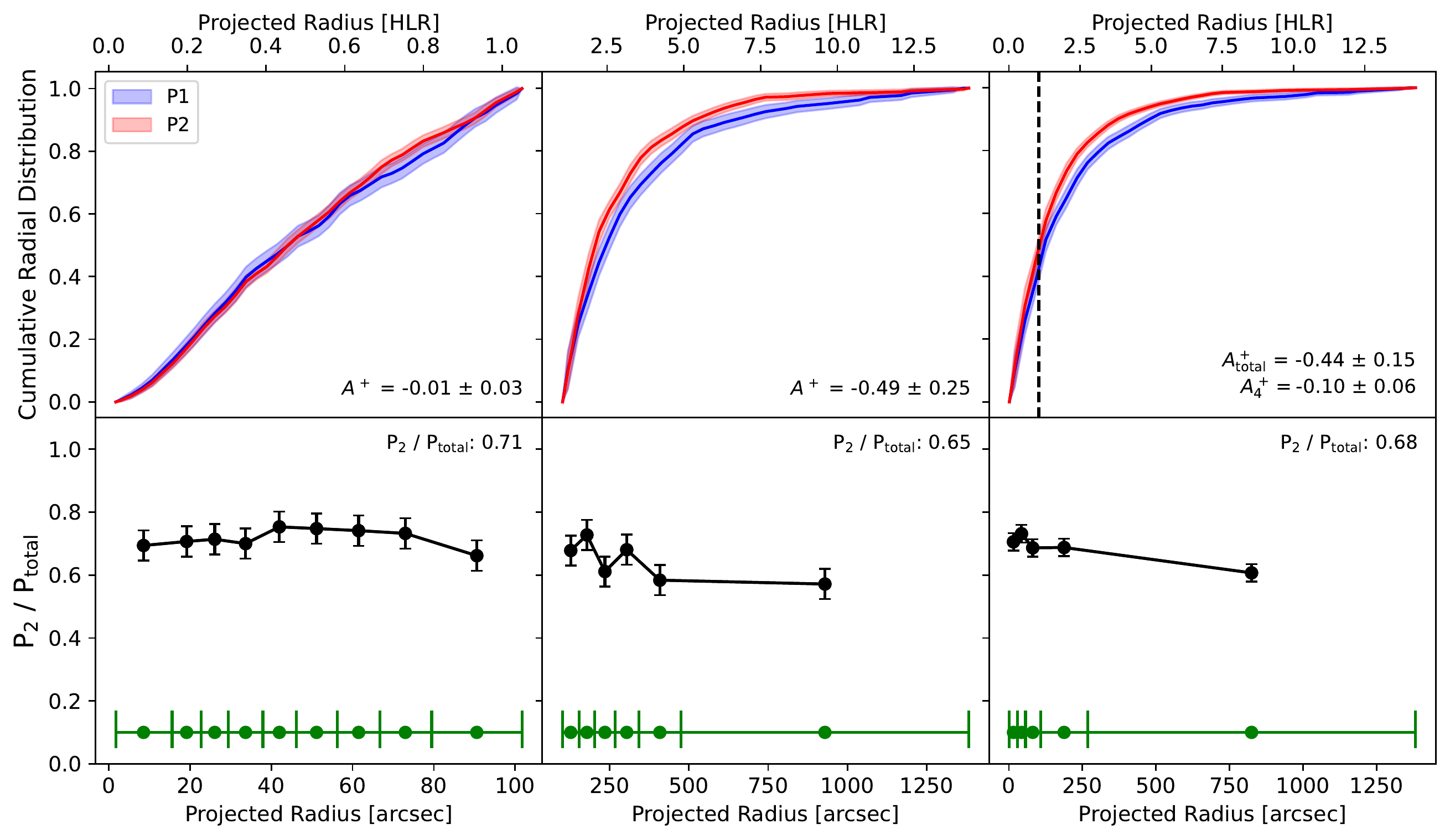}
    \caption{As in Figure \ref{fig:Aplus}, but for NGC 5904.}
    \label{fig:Aplus5904}
\end{figure*}

\begin{figure*}
    \centering
    \includegraphics[width=\textwidth]{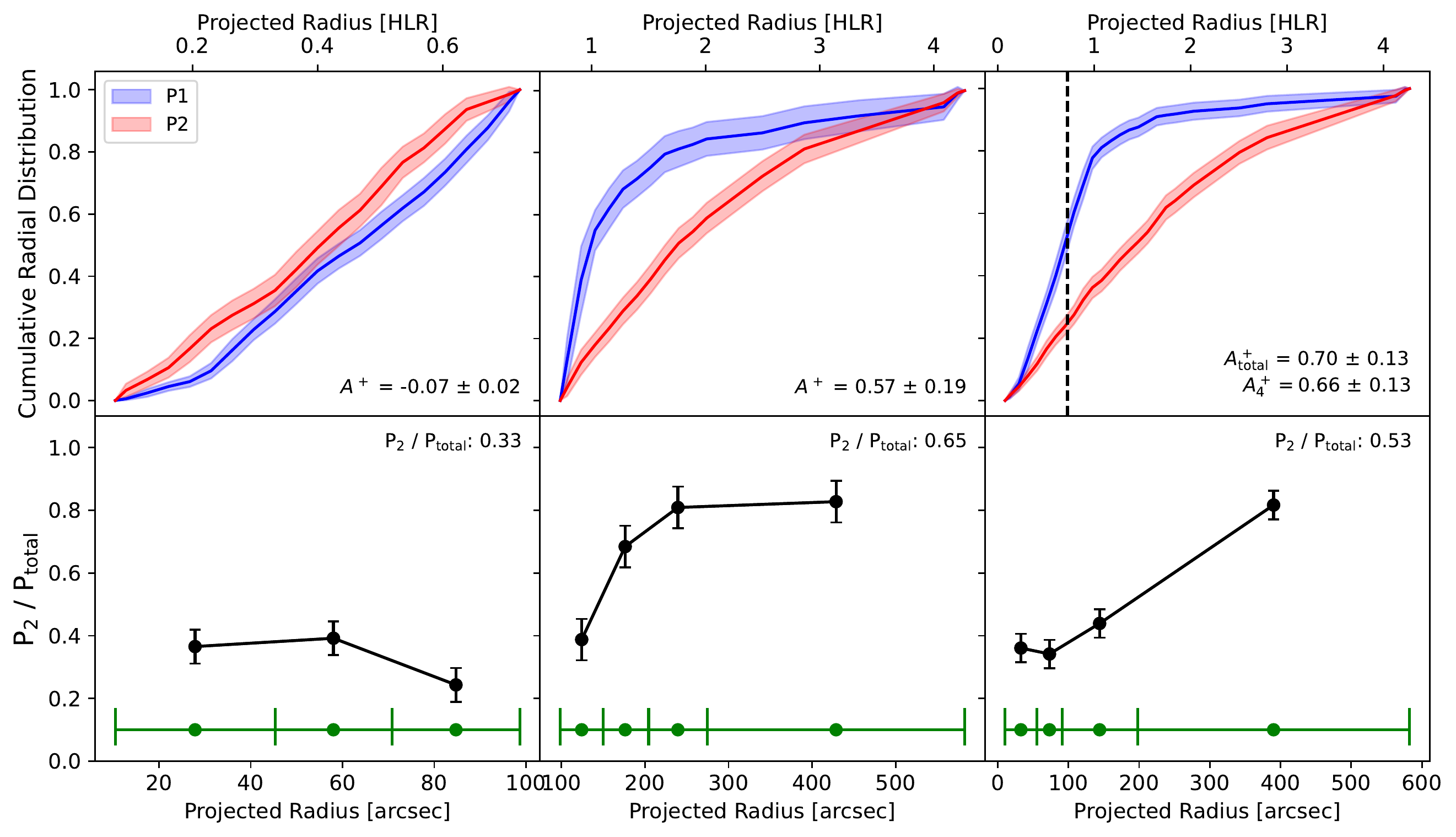}
    \caption{As in Figure \ref{fig:Aplus}, but for NGC 6101.}
    \label{fig:Aplus6101}
\end{figure*}

\begin{figure*}
    \centering
    \includegraphics[width=\textwidth]{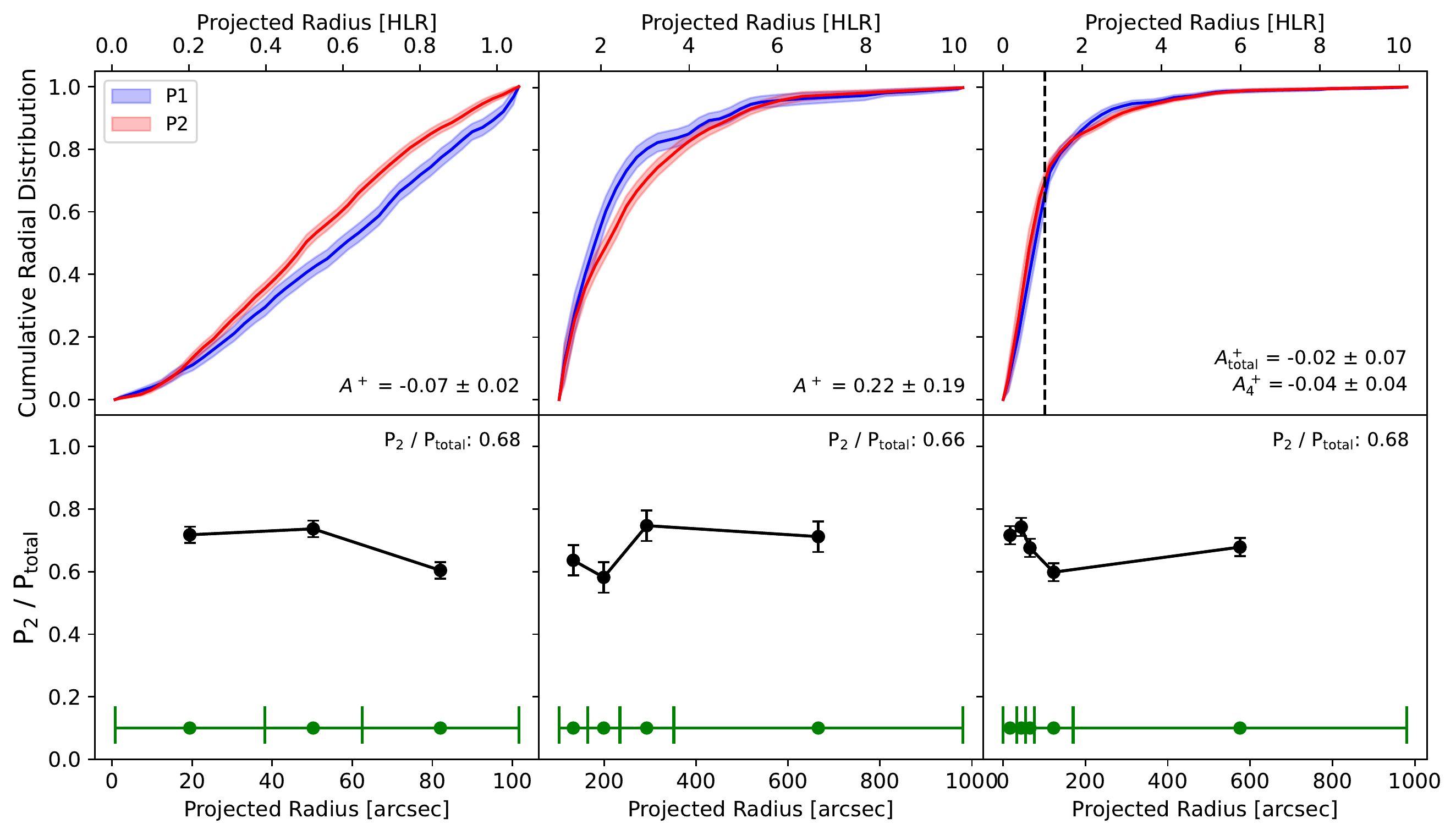}
    \caption{As in Figure \ref{fig:Aplus}, but for NGC 6205.}
    \label{fig:Aplus6205}
\end{figure*}

\begin{figure*}
    \centering
    \includegraphics[width=\textwidth]{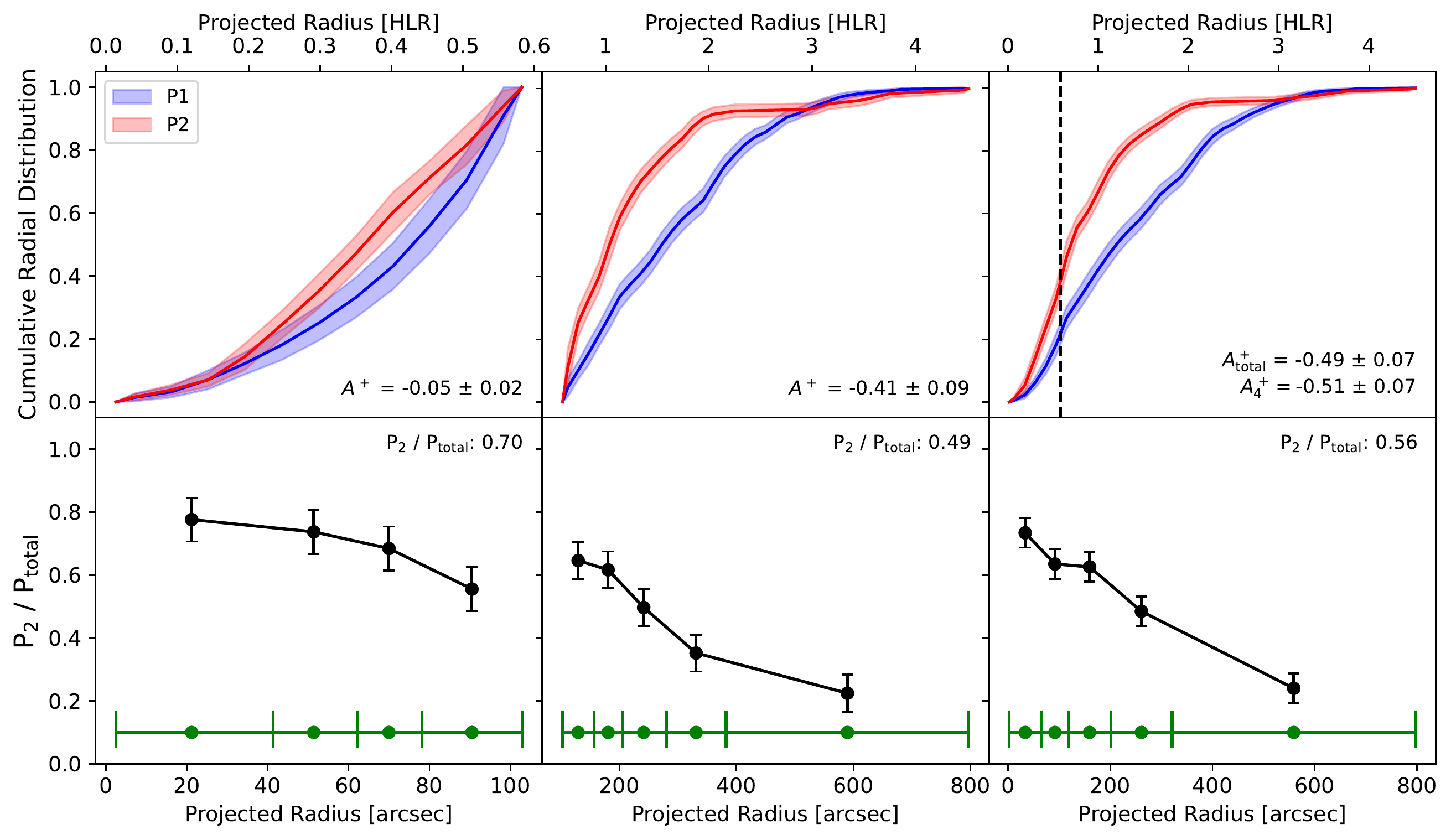}
    \caption{As in Figure \ref{fig:Aplus}, but for NGC 6809.}
    \label{fig:Aplus6809}
\end{figure*}

\begin{figure*}
    \centering
    \includegraphics[width=\textwidth]{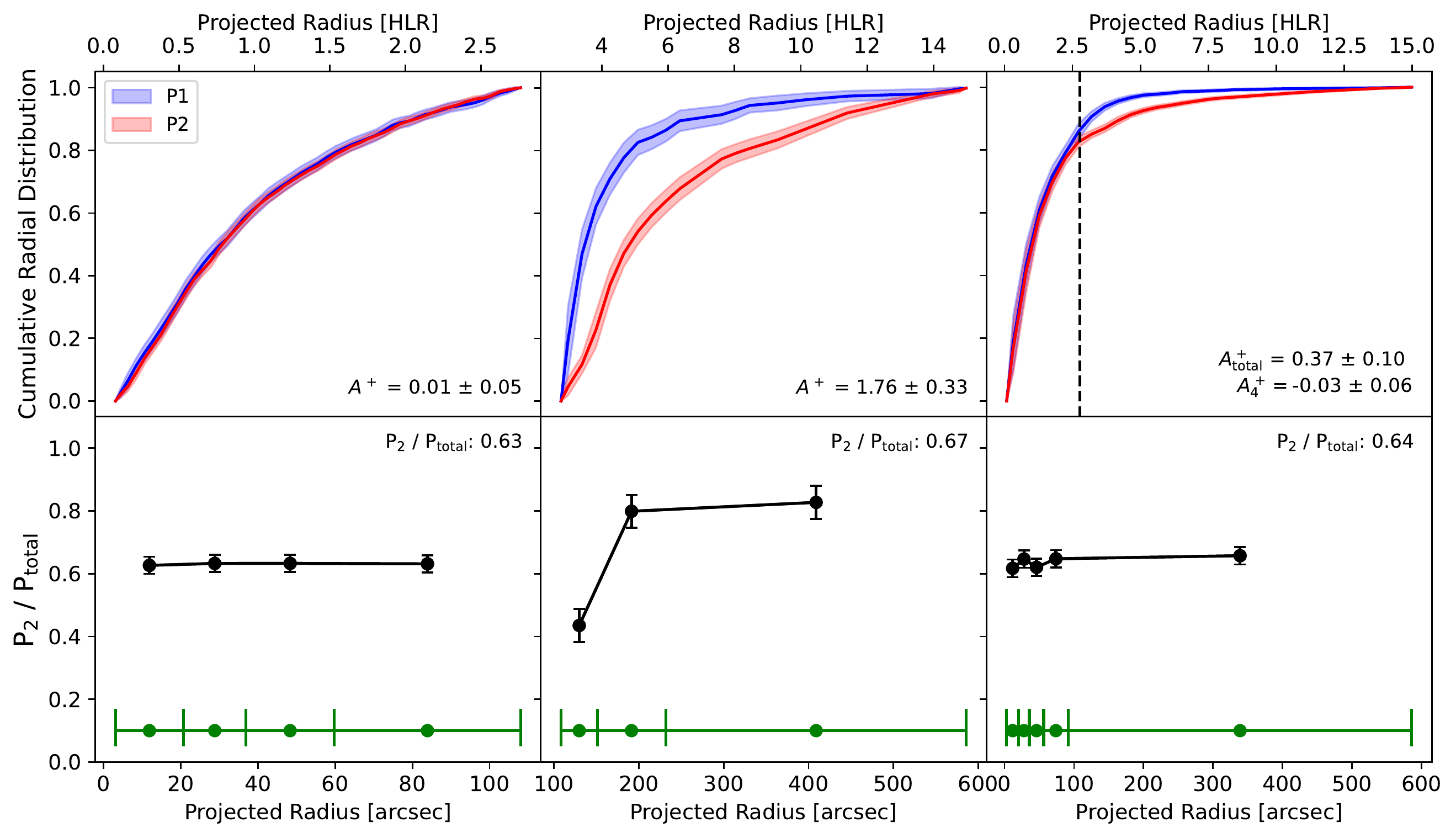}
    \caption{As in Figure \ref{fig:Aplus}, but for NGC 7078. Refer to Section \ref{sec:Apluszero} for a special discussion on NGC 7078.}
    \label{fig:Aplus7078}
\end{figure*}

\begin{figure*}
    \centering
    \includegraphics[width=\textwidth]{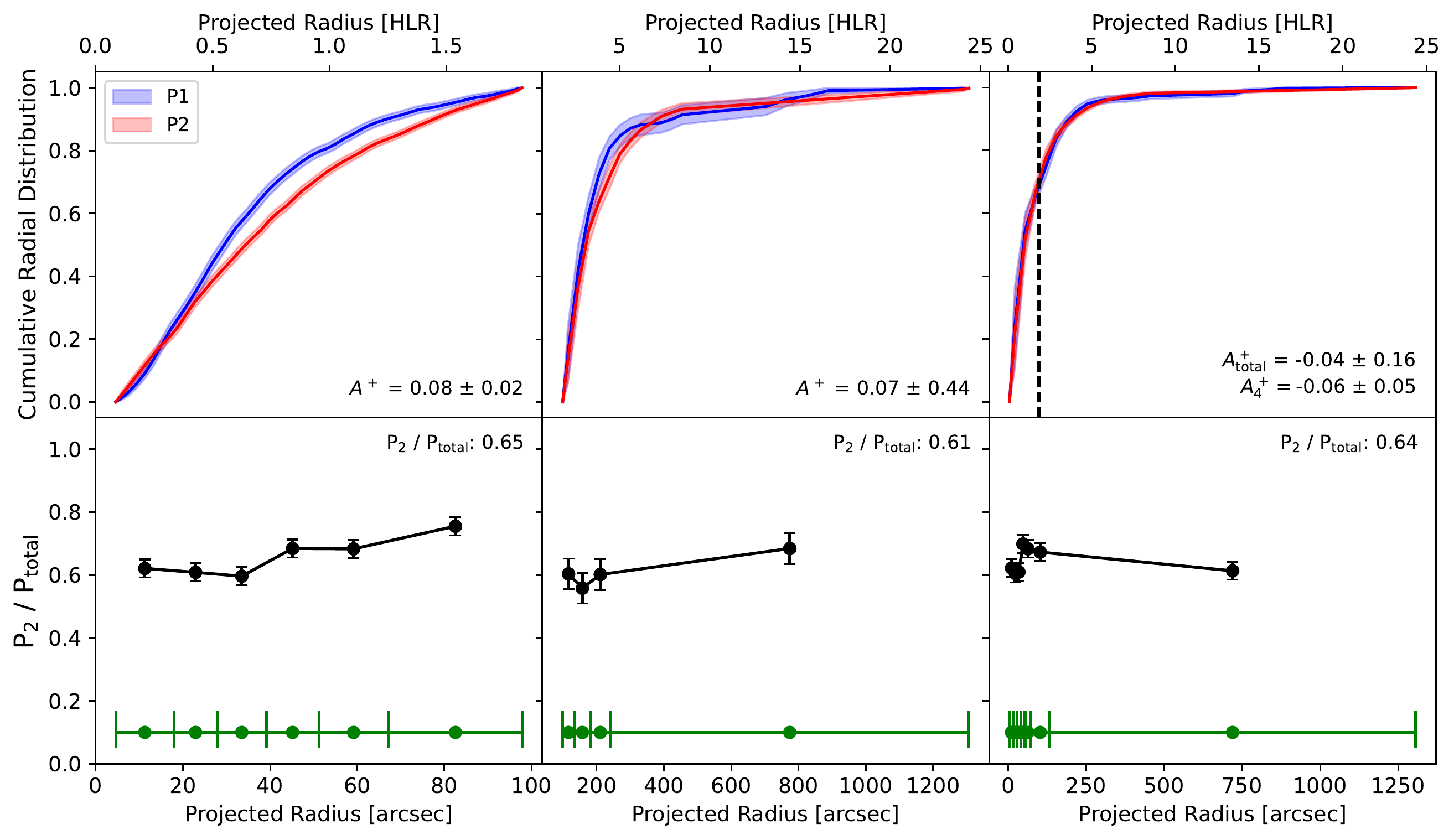}
    \caption{As in Figure \ref{fig:Aplus}, but for NGC 7089.}
    \label{fig:Aplus7089}
\end{figure*}


\bsp	
\label{lastpage}
\end{document}